\documentclass[aps,pre,preprint,groupedaddress,showpacs]{revtex4-1}
\usepackage{bm}
\usepackage{amsmath}
\usepackage{epsfig}
\usepackage{longtable}
\usepackage{graphicx}

\def\gazz{\mathrel{\mathchoice {\vcenter{\offinterlineskip\halign{\hfil
$\displaystyle##$\hfil\cr>\cr\sim\cr}}}
{\vcenter{\offinterlineskip\halign{\hfil$\textstyle##$\hfil\cr>\cr\sim\cr}}}
{\vcenter{\offinterlineskip\halign{
\hfil$\scriptstyle##$\hfil\cr>\cr\sim\cr}}}
{\vcenter{\offinterlineskip\halign{\hfil$\scriptscriptstyle##
$\hfil\cr>\cr\sim\cr}}}}}		
\def\pr{\prime}
\def\be{\begin{equation}}
\def\lan{\left\langle}
\def\ran{\right\rangle}
\def\ee{\end{equation}}
\def\barr{\begin{array}}
\def\earr{\end{array}}

\def\l{\left}
\def\r{\right}
\def\dis{\displaystyle}
\def\ed{\end{document}}

\def\bee{{\mbox{\boldmath $E$}}}

\def\cg{{\cal G}}
\def\ce{{\cal E}}
\def\cm{{\bf m}}
\def\cs{{\bf s}}
\def\ct{{\bf t}}

\def\spin{\frac{1}{2}}
\def\wh{{\widehat {H}}}
\def\whh{{\widehat {h}}}
\def\wv{{\widehat {V}}}
\def\cad{{\cal D}}
\def\cads{{\mbox{\boldmath $d$}}}
\def\wm{{\widehat {M}}}
\def\we{{\widehat {E}}}
\def\cS{{\cal S}}
\def\dg{{\dagger}}

\begin{document}

\title{Spectral properties of two-body random matrix ensembles for boson
systems with spin}

\author{Manan Vyas$^{1}$, N.D. Chavda$^{2}$,  V.K.B.
Kota$^{1,3,}$\footnote{Corresponding author, phone: +91-79-26314464, Fax:
+91-79-26314460 \\ {\it E-mail address:}  vkbkota@prl.res.in (V.K.B. Kota)}
and  V. Potbhare$^{2}$}

\affiliation{$^1$Physical Research Laboratory, Ahmedabad 380 009, India\\
$^2$Applied Physics Department, Faculty of Technology and Engineering, M. S.
University of Baroda, Vadodara 390 001, India\\
$^3$Department of Physics, Laurentian University, Sudbury,
Ontario, Canada  P3E 2C6}

\begin{abstract}

For $m$ number of bosons, carrying spin ($\cs=\spin$) degree of freedom, in
$\Omega$ number of single particle orbitals, each doubly degenerate, we
introduce and analyze embedded Gaussian orthogonal ensemble of random
matrices generated by random two-body interactions that are spin ($S$)
scalar [BEGOE(2)-$\cs$]. Embedding algebra for the BEGOE(2)-$\cs$ ensemble
and also for BEGOE(1+2)-$\cs$ that includes the mean-field one-body part  is
$U(2\Omega) \supset U(\Omega) \otimes SU(2)$ with $SU(2)$ generating spin. 
A method for constructing the ensembles in fixed-($m,S$) spaces has been
developed. Numerical calculations show that for BEGOE(2)-$\cs$, the 
fixed-$(m,S)$ density of states is close to Gaussian and level fluctuations
follow GOE in the  dense limit. For BEGOE(1+2)-$\cs$, generically there is
Poisson to GOE transition in level fluctuations as the interaction
strength (measured in the units of the average spacing of the single
particle levels defining the mean-field) is  increased. The interaction
strength needed for the onset of the transition  is found to decrease with
increasing $S$. Propagation formulas for the fixed-$(m,S)$ space energy
centroids and spectral variances are derived for a general one plus two-body
Hamiltonian preserving spin. Derived also is the formula for the variance
propagator for the fixed-$(m,S)$ ensemble averaged spectral variances. Using
these, covariances in energy centroids and spectral variances are analyzed. 
Variance propagator clearly shows, by applying the Jacquod and Stone
prescription, that the BEGOE(2)-$\cs$ ensemble generates ground states with
spin $S=S_{max}$. This is further corroborated by analyzing the structure of
the ground states in the  presence of the exchange interaction $\hat{S}^2$
in BEGOE(1+2)-$\cs$.  Natural spin ordering ($S_{max}$, $S_{max}-1$,
$S_{max}-2$, $\ldots$, $0$ or $\spin$) is also observed with random
interactions. Going beyond these,  we also introduce pairing symmetry in the
space defined by BEGOE(2)-$\cs$. Expectation values of the pairing
Hamiltonian show that random interactions exhibit pairing correlations in
the ground state region. 

\end{abstract}


\maketitle

\section{Introduction}

Random matrix theory  has been established to be one of the central themes
for chaotic quantum systems \cite{Ha-10}.  The classical Gaussian orthogonal
(GOE), unitary (GUE) and symplectic (GSE) ensembles of random matrices are
ensembles of multi-body, not two-body interactions. However, finite
interacting quantum systems, such as nuclei, atoms, quantum dots, small
metallic grains and interacting spin systems modeling quantum computing core
and BEC, are governed largely by two-body interactions and hence, it is
important to consider ensembles generated by random two-body interactions.
These ensembles are defined by representing the two-particle Hamiltonian by
one of the classical ensembles (GOE, GUE, GSE) and then the many particle
($m>2$) Hamiltonian is generated by exploiting  the direct product structure
of the $m$-particle Hilbert spaces. As a random matrix ensemble in the
two-particle spaces is embedded in the many particle Hamiltonian, these
ensembles are generically called embedded ensembles (EE) and with GOE
embedding, they will be EGOE's. A wide variety of EE for fermions have been
introduced in literature \cite{Fr-70,Bo-71,MF-75,Ko-01,Ma-10,Go-10}. 

Simplest of EE is the embedded Gaussian orthogonal ensemble for spinless
fermion systems generated by random two-body interactions, denoted by
EGOE(2). In general, it is also possible to define EGOE($k$) ensembles 
generated by $k$-body ($k<m$) interactions \cite{MF-75}.  It is useful to
mention that many diversified methods like numerical Monte-Carlo methods,
binary correlation approximation, trace propagation, group theory,
supersymmetry and perturbation theory are used to derive generic properties
of EE \cite{MF-75,Be-01,Ko-05,Ko-07,Ma-10b}.  Some of the generic results
for EGOE($k$) are as follows: (i) eigenvalue density exhibits, with
increasing $m$, transition from semicircle to Gaussian with $m=2k$ being the
transition point \cite{Be-01}; (ii) numerical studies have shown that the
level and strength  fluctuations follow GOE \cite{Br-81};  (iii) there is
average-fluctuation separation with increasing $m$ \cite{Br-81,Le-08};  (iv)
ensemble averaged transition strength densities follow bivariate Gaussian
form and consequently, transition strength sums will be close to a ratio of
two Gaussians \cite{FKPT}; and (v) there will be non-zero  correlations
between states with different particle numbers \cite{PW-07,Ko-06}. Besides
two-body interactions, Hamiltonians for realistic systems contain a
mean-field part [defined by non-degenerate single particle (sp) levels] and
then the ensemble is  denoted by EE(1+2). This ensemble exhibits,  with
increasing strength $\lambda$ of the interaction ($\lambda$ is in units of
the average sp level spacing), three chaos markers defining transitions in
level fluctuations ($\lambda_c$), strength functions ($\lambda_F$) and
entropy ($\lambda_d$) respectively. Generic results of EE(2) are valid for
EE(1+2) in the strong coupling regime ($\lambda >> \lambda_F$). Realistic
systems also preserve various symmetries. For example, spin $S$ is a good
quantum number for atoms and quantum dots, angular momentum $J$ and parity
$\pi$ are good quantum numbers for nuclei and so on. Therefore, it is more
appropriate to study EE with good symmetries. A simple but non-trivial
extension of EE  is to consider finite quantum systems with spin
($\cs=\spin$) and then the ensembles are called  EE(1+2)-$\cs$. For finite
interacting fermion systems, EGOE(1+2)-$\cs$ has been studied in detail
using a mixture of numerical and analytical techniques
\cite{Ko-06a,Ma-09,Ma-10} and EGUE(2)-$\cs$ using Wigner-Racah algebra
\cite{Ko-07}. Moreover, EGUE(2) with spin-isospin $SU(4)$ symmetry,
comparatively complicated than the random matrix models analyzed before, has
been recently introduced and analyzed in some detail \cite{Ma-10b}. More
importantly, there are now several applications of EE to mesoscopic systems
\cite{Meso1,Meso2,JS-01,Ma-10c}, quantum information science \cite{Br-08}
and in investigating  thermalization in finite quantum systems
\cite{Ri-09,Sa-10,Sa-10a,Ol-09}. Unlike for fermion systems, there are only
a few EE investigations for finite interacting boson systems
\cite{PDPK,Ag-01,Ag-02,Ch-03,Ch-04}; the corresponding EE are called BEE (B
stands for bosons). Briefly, these studies are as follows.

Firstly, it is important to mention that, unlike fermion systems,  for
interacting spinless boson systems with $m$ bosons in $N$ sp orbitals, 
dense limit defined by $m \to \infty$, $N  \to \infty$ and $m/N  \to \infty$
is also possible as $m$ can be greater than $N$ for bosons. It is now well
understood that BEGOE(2) [also BEGUE(2)] generates in  the dense limit,
eigenvalue density close to a Gaussian \cite{KP-80,PDPK}. Also the ergodic
property is found to be valid in the dense limit with sufficiently large $N$
\cite{Ch-03}; there are deviations for small $N$ \cite{Ag-01}. Similarly,
for BEGOE(1+2),  as the strength $\lambda$ of the two-body interaction
increases, there is Poisson to GOE transition in  level fluctuations at
$\lambda=\lambda_c$ \cite{Ch-03} and with further increase in $\lambda$,
there is Breit-Wigner to Gaussian transition in strength functions
\cite{Ch-04}. For BEGUE($k$), exact analytical results for the lowest two
moments of the two-point function have been derived by Agasa et al
\cite{Ag-02}.  Level fluctuations and wavefunction structure in interacting 
boson systems are also studied using interacting boson models of atomic
nuclei \cite{Al-91,Al-93,Ca-00} and a  symmetrized two coupled rotors model
\cite{Bo-98a,Bo-98b}. In addition, using random interactions in interacting
boson models, there are several studies on the generation of regular
structures in boson systems with random interactions
\cite{Ku-97,BF-00,Ko-04,Yo-09}. Finally, there are also studies on
thermalization in finite quantum systems using boson systems
\cite{Ri-09,Sa-10,Sa-10a,Ol-09}.

Going beyond the embedded ensembles for spinless boson systems, our purpose
in this paper is to introduce and analyze spectral properties of embedded
Gaussian orthogonal ensemble of random matrices for boson systems with spin
degree of freedom [BEGOE(2)-$\cs$ and also BEGOE(1+2)-$\cs$] and for
Hamiltonians that conserve the total spin of the $m$-boson systems. Here the
spin is, for example, as the $F$-spin in the proton-neutron interacting
boson model ($pn$IBM) of atomic nuclei \cite{Ca-05}. Just as the earlier
embedded ensemble studies \cite{Ag-01,Ag-02,Ch-03,Ch-04}, a major motivation
for the study undertaken in the present paper is its possible applications
to ultracold atoms. There are several studies of the properties of a mixture
of two species of atoms which correspond to pseudospin-$\spin$ bosons (i.e.
two-component boson systems)  with $m_\cs=\pm \spin$ distinguishing the two
species; see for example \cite{Al-03,Yu-10}.  However, the Hamiltonians
appropriate for these studies do not conserve the total spin (as the system
does not have true $\spin$-spins) and therefore,  the model study presented
in the present paper will not be directly applicable to these systems in
understanding their statistical properties. Nevertheless, the
BEGOE(1+2)-$\cs$ with spin-$\spin$ bosons is a simple yet non-trivial
extension of the spinless BEE. This ensemble is useful in obtaining  several
physical conclusions, like spin dependence of the order to chaos transition
marker in level fluctuations, the spin of the ground state (gs), the spin
ordering of excited states and pairing correlations in the gs region
generated by random interactions, that explicitly require inclusion of spin
degree of freedom (these are discussed in Sections III, V and VI). It should
be emphasized that the present paper opens a new direction in defining and
analyzing embedded ensembles for boson systems with symmetries. It is
important to mention here  that there are now many studies of spinor BEC
using Hamiltonians conserving the total spin with the bosons carrying
$\cs=1$ (also higher) degree of freedom \cite{Pe-10,Yi-07}. Extensions of
BEGOE(1+2)-$\cs$ with $\cs=\spin$ to boson ensembles  with integer spin
$\cs=1$ (or higher) is for future. Now we will give a preview.

In Section II, introduced is the new embedded ensemble BEGOE(2)-$\cs$ [and
also BEGOE(1+2)-$\cs$] for a system of $m$ bosons in $\Omega$ number of sp
orbitals that are doubly degenerate with total spin $S$ being a good
symmetry. A method for the numerical construction of this ensemble in
fixed-$(m,S)$ spaces is  described. Numerical results for the ensemble
averaged  eigenvalue density, nearest neighbor spacing distribution and the
long-range rigidity measure $\overline{\Delta}_3$ are presented in Section
III. Propagation formulas for fixed-$(m,S)$ energy centroids and spectral
variances for general one plus two-body Hamiltonians that preserve $S$ are
given in Section IV. Here, given also is the analytical  formula for the
ensemble averaged fixed-$(m,S)$  spectral variances. Using these, studied
are  covariances in energy centroids and spectral variances generated by
BEGOE(2)-$\cs$ ensemble between states with different ($m$, $S$). Section V
gives results for the preponderance of maximum $S$-spin ground states and
natural spin order generated by random interactions. Here, exchange
interaction is added to the BEGOE(1+2)-$\cs$ Hamiltonian. Pairing in
BEGOE(2)-$\cs$ is introduced in Section VI and presented also are some
numerical results for pairing correlations. Finally, Section VII gives
conclusions and future outlook.

\section{Definition and construction of BEGOE(1+2)-$\cs$} 

Let us consider a system of $m$ ($m>2$) bosons distributed in $\Omega$
number of sp orbitals each with spin $\cs=\spin$. Then the number of sp 
states is $N=2\Omega$. The sp  states are denoted by  $\l.\l| i,m_\cs=\pm
\spin\r.\ran$ with $i=1,2,\ldots,\Omega$ and the two particle symmetric
states are denoted by $\l.\l|(ij)s,m_s\r.\ran$ with $s=0$ or $1$. It is
important to note that for EGOE(1+2)-$\cs$, the embedding algebra is
$U(2\Omega) \supset U(\Omega) \otimes SU(2)$ with $SU(2)$ generating spin;
see Sections V and VI ahead.  The dimensionalities of the two-particle
spaces with $s=0$  and $s=1$ are $\Omega(\Omega-1)/2$ and
$\Omega(\Omega+1)/2$ respectively. For one plus two-body Hamiltonians
preserving $m$ particle spin $S$, the one-body Hamiltonian is
$\whh(1)=\sum_{i=1}^\Omega\, \epsilon_i n_i$ where the orbitals $i$ are
doubly degenerate,  $n_i$ are number operators  and $\epsilon_i$ are sp
energies (it is in principle possible to consider $\whh(1)$ with
off-diagonal energies $\epsilon_{ij}$).  The two-body Hamiltonian $\wv(2)$
preserving $m$ particle spin $S$ is defined by the symmetrized  two-body
matrix elements  $V^s_{ijkl}=\lan (kl)s,m_s \mid \wv(2) \mid (ij)s,m_s\ran$
with $s=0,\,1$ and they are independent of the $m_s$ quantum number; note
that for $s=0$, only $i \neq j$ and $k \neq l$ matrix elements exist. Thus
$\wv(2)=\wv^{s=0}(2) + \wv^{s=1}(2)$  and the sum here is a direct sum. The
BEGOE(2)-$\cs$ ensemble for a given $(m,S)$ system is generated by first  
defining the two parts of the two-body Hamiltonian to be independent GOEs in
the two-particle spaces [one for $\wv^{s=0}(2)$ and other for
$\wv^{s=1}(2)$], with the matrix elements variances being unity (except for
diagonal matrix elements whose variance is two). Now the $V(2)$ ensemble
defined by  $\{\wv(2)\}=\{\wv^{s=0}(2)\} + \{\wv^{s=1}(2)\}$ is propagated
to the  $(m,S)$-spaces by using the geometry (direct product structure) of
the $m$-particle spaces; here $\{\;\}$ denotes ensemble. By adding the
$\whh(1)$ part, the BEGOE(1+2)-$\cs$ is defined by the operator
\be
\{\wh\}_{\mbox{BEGOE(1+2)-\cs}} = \whh(1)+  \lambda_0\, \{\wv^{s=0}(2)\} +
\lambda_1\, \{\wv^{s=1}(2)\}\,.
\label{eq.begoe-s}
\ee
Here $\lambda_0$ and $\lambda_1$ are the strengths of the $s=0$ and  $s=1$
parts of $\wv(2)$ respectively.  The mean-field one-body Hamiltonian
$\whh(1)$ in Eq. (\ref{eq.begoe-s}) is defined by sp energies $\epsilon_i$
with average spacing $\Delta$. Without loss  of generality, we put
$\Delta=1$ so that $\lambda_0$ and $\lambda_1$ are in the  units of
$\Delta$.  In the present paper, we choose sp energies $\epsilon_i=i+1/i$ as
in our previous studies of fermion systems \cite{Ma-10}. In principle, many
other choices for the sp energies are possible.  Thus BEGOE(1+2)-$\cs$ is
defined by the five  parameters $(\Omega, m, S, \lambda_0, \lambda_1)$. The
$H$ matrix dimension $d(\Omega,m,S)$ for a given $(m,S)$ is
\be
d(\Omega,m,S)=\frac{(2S+1)}{(\Omega-1)} { \Omega+m/2+S-1
\choose m/2+S+1} {\Omega+m/2-S-2 \choose m/2-S}\;,
\label{eq.msdim}
\ee
and they satisfy the sum rule $\sum_S\;(2S+1)\;d(\Omega,m,S)= {N+m-1 \choose
m}$.  For example: (i)   $d(4,10,S)=196$, $540$, $750$,  $770$, $594$ and
$286$ for spins $S=0-5$; (ii)  $d(4,11,S)=504$, $900$, $1100$, $1056$, $780$
and $364$ for $S=1/2-11/2$; (iii)   $d(5,10,S)=1176$,  $3150$, $4125$,
$3850$, $2574$ and $1001$ for $S=0-5$; (iv) $d(6,12,S)=13860$, $37422$,
$50050$, $49049$, $36855$, $20020$ and $6188$ for $S=0-6$; and (v) 
$d(6,16,S)=70785$, $198198$, $286650$, $321048$, $299880$, $235620$,
$151164$, $72675$ and $20349$ for $S=0-8$.

Given $\epsilon_i$ and $V^s_{ijkl}$, the many particle Hamiltonian matrix
for a given ($m,S$) can be constructed using the $M_S$ representation ($M_S$
is the $S_z$ quantum number) and for spin projection the $S^2$ operator is
used as it was done for fermion systems in \cite{Ko-06a}. Alternatively, it
is possible to construct the $H$ matrix directly in a good $S$ basis using
angular-momentum algebra as it was done for fermion systems in
\cite{Tu-06}.  We have employed the $M_S$ representation for constructing
the $H$ matrices with $M_S=M_S^{min}=0$ for even $m$ and
$M_S=M_S^{min}=\spin$ for odd $m$ and they will contain states with all $S$
values.  The dimension of this basis space is $\cad(\Omega,m,M_S^{min}) =
\sum_S\,d(\Omega,m,S)$.  For example, $\cad(4,10,0)=3136$,
$\cad(4,11,\spin)=4704$,   $\cad(5,10,0)=15876$, $\cad(6,12,0)=213444$ and
$\cad(6,16,0)=1656369$. 

To construct the many particle Hamiltonian matrix for a given $(m,S)$, 
first the  sp states $\l.\l| i,m_\cs=\pm \spin\r.\ran$ are arranged in such
a way that the first $\Omega$ states have $m_\cs=\spin$ and the remaining
$\Omega$ states have $m_\cs=-\spin$ so that the sp states are
$\l.\l|r\r.\ran=\l.\l|i=r,m_\cs=\spin\r.\ran$ for $r \leq \Omega$ and
$\l.\l| r\r.\ran=\l.\l|i=r-\Omega,m_\cs=-\spin\r.\ran$ for $r > \Omega$.
Using the direct product structure of the many-particle states,  the
$m$-particle configurations $\cm$, in occupation number representation, are
\be
\cm=\left| {\prod\limits_{r=1}^{N=2\Omega} {m_r } } \right\rangle  =
\left| {m_1, m_2, \ldots, m_\Omega, m_{\Omega+1}, m_{\Omega+2},
\ldots, m_{2\Omega} } \right\rangle,
\label{eq.ms-conf}
\ee
where $m_r \geq 0$ with $\sum_{r=1}^N{m_r}=m$ and 
$M_S=\spin\l[\sum_{r=1}^{\Omega}\,m_r - \sum_{r^\prime=
\Omega+1}^{2\Omega}\,m_{r^\prime}\r]$. To proceed further, the (1+2)-body
Hamiltonian defined by $\epsilon_i$ and $V^{s=0,1}_{ijkl}$ is converted into
the $\l.\l| i,m_\cs=\pm \spin\r.\ran$ basis. Then the sp energies 
$\epsilon_i^\pr$ with $i=1,2,\ldots,N$ are $\epsilon_i^\pr = 
\epsilon_{i+\Omega}^\pr = \epsilon_i$ for $i \leq \Omega$. Similarly,
$V^{s}_{ijkl}$ are changed to  $V_{im_i,jm_j,km_k,lm_l}={\left\langle {im_i,
jm_j } \right|\left. {V(2) } \right|\left. {km_k, lm_l} \right\rangle }$
using,
\be
\barr{rcl}
V_{i\spin,j\spin,k\spin,l\spin} & = & 
V_{i -\spin,j -\spin,k -\spin,l -\spin} = 
V^{s=1}_{ijkl} \;,\\ \\
V_{i\spin,j -\spin,k\spin,l -\spin} & = & \dis\frac{\sqrt{(1+\delta_{ij})
(1+\delta_{kl})}}{2} \l[ V^{s=1}_{ijkl}+ V^{s=0}_{ijkl}\r]\;,
\earr \label{tbme_new}
\ee
with all the other matrix elements being zero except for the symmetries, 
\be 
V_{im_i,jm_j,km_k,lm_l} = V_{km_k,lm_l,im_i,jm_j}=
V_{jm_j,im_i,lm_l,km_k} = V_{im_i,jm_j,lm_l,km_k}\;. 
\ee 
Using $(\epsilon_r^\pr, V_{im_i,jm_j,km_k,lm_l})$'s, construction of the 
$m$-particle $H$ matrix in the basis defined by Eq. (\ref{eq.ms-conf}) 
reduces to the problem of BEGOE(1+2) for spinless boson systems and hence
Eq. (4) of \cite{PDPK} will give the formulas for the non-zero matrix
elements. For completeness, these formulas are given in Appendix A. Now
diagonalizing the $S^2$ matrix in the basis defined by  Eq.
(\ref{eq.ms-conf}) will give the unitary transformation required to  change
the $H$ matrix in $M_S$ basis into good $S$ basis.  Following this method,
we have numerically constructed BEGOE(1+2)-$\cs$ in many examples and
analyzed various spectral properties generated by this ensemble.  In
addition, we have also derived some analytical results as discussed ahead in
Sections IV and VI.  These results are also used to validate the
BEGOE(1+2)-$\cs$ numerical code we have developed. In addition, we have also
verified the code by directly programming the operations that give Eq.
(\ref{eq.app2}). In this paper, we deal mainly  with BEGOE(2)-$\cs$ and the
focus is on the dense limit defined by   $m \to \infty$, $\Omega \to
\infty$, $m/\Omega  \to \infty$ and $S$ is  fixed. Now we will discuss these
results.

\section{Numerical results for eigenvalue density and level fluctuations in
the dense limit}

\begin{figure}
\includegraphics[width=5in,height=6.5in]{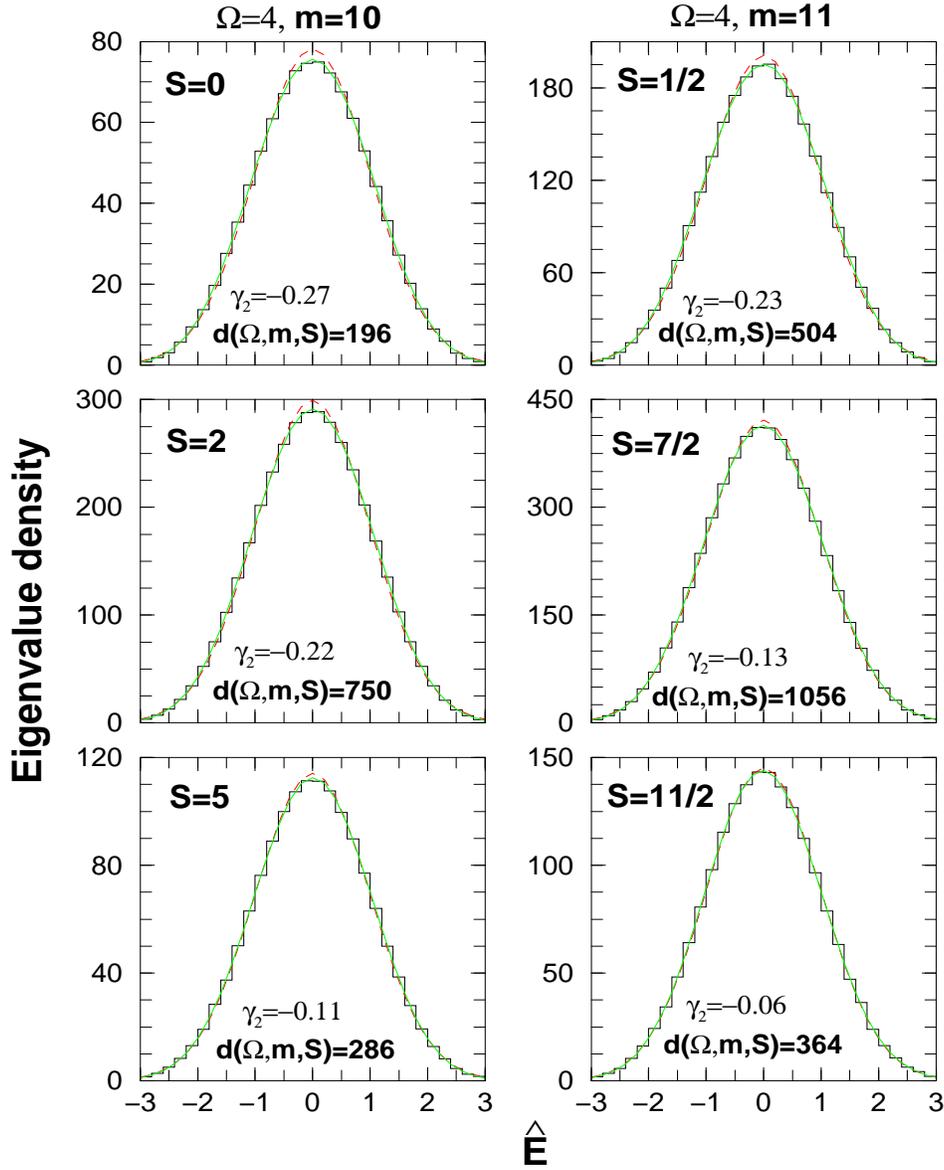}
\caption{(Color online) Ensemble averaged eigenvalue density 
$\rho^{m,S}(\widehat{E})$ vs $\widehat{E}$ for BEGOE(2)-$\cs$ ensembles with
$\Omega=4$, $m=10$ and $\Omega=4$, $m=11$.  In the figure, histograms
constructed with a bin size $0.2$  are BEGOE(2)-$\cs$ results and they are
compared with Gaussian (dashed red) and Edgeworth (ED) corrected Gaussian
(solid green) forms.  The ensemble averaged values of the excess  parameter
$(\gamma_2)$ are also shown in the figure.  In the plots, the area under the
curves is normalized to the dimensions $d(\Omega,m,S)$. See text for further
details.}
\label{den}
\end{figure}

We begin with the ensemble averaged  fixed-($m$, $S$) eigenvalue density
$\rho^{m,S}(E)$, the one-point function for eigenvalues. First we present
the results for BEGOE(2)-$\cs$ ensemble defined by $\whh(1)=0$  
in Eq. (\ref{eq.begoe-s}) and then the Hamiltonian operator is,
\be
\{\wh\}_{\mbox{BEGOE(2)-\cs}} =  \lambda_0\, \{\wv^{s=0}(2)\} +
\lambda_1\, \{\wv^{s=1}(2)\}\,.
\label{eq.v2}
\ee
We have considered a 500 member BEGOE(2)-$\cs$ ensemble with $\Omega=4$ and 
$m=10$ and similarly a 100 member ensemble with $\Omega=4$ and  $m=11$. 
Here and in all other numerical results presented in the paper, we use
$\lambda_0 = \lambda_1 = \lambda$. In the construction of the ensemble
averaged  eigenvalue densities, the spectra of each member of the ensemble
is first zero centered and scaled to unit width (therefore the densities are
independent of the $\lambda$ parameter).  The eigenvalues are then denoted
by $\widehat{E}$. Given the fixed-($m,S$) energy centroids $E_c(m,S)$ and
spectral widths $\sigma(m,S)$, $\widehat{E}=[E-E_c(m,S)]/\sigma(m,S)$.  Then
the histograms  for the density are generated by combining the eigenvalues
$\widehat{E}$ from all the members of the  ensemble. Results are shown in
Fig. \ref{den} for a few selected $S$ values. The calculations have been
carried out for all $S$ values (the results for other $S$ values are close
to those given in the figure) and also for many other BEGOE(2)-$\cs$
examples.  It is clearly seen that the eigenvalue densities are close to
Gaussian (denoted by $\cg$ below) with the ensemble averaged skewness
($\gamma_1$) and excess ($\gamma_2$) being very small; $|\gamma_1| \sim 0$,
$|\gamma_2| \sim 0.1-0.27$.  The agreements with Edgeworth (ED) corrected
Gaussians are excellent. The ED form that includes $\gamma_1$ and $\gamma_2$
corrections is given by
\be
\barr{rcl}
\eta_{ED}(\we) & = &
\eta_{\cg}(\we)
\l\{1+\l[{\dis\frac{\gamma_1}{6}}He_3\l(\we\r)\r]+
\l[{\dis\frac{\gamma_2}{24}}He_4\l(\we\r) +
{\dis\frac{\gamma_1^2}{72}} He_6\l(\we\r) \r]\r\}\;;\\
\eta_{\cg}(\we) & = &
\dis\frac{1}{\sqrt{2\pi}}\exp
\l(-\dis\frac{\we^2}{2}\r) \;.
\earr \label{eq.gau1}
\ee
Here, $He$ are Hermite polynomials: $He_3(x)=x^3-3x$, $He_4(x)=x^4-6x^2+3$
and $He_6(x)= x^6-15x^4 + 45x^2-15$. 

\begin{figure}
\includegraphics[width=5.5in,height=6.75in]{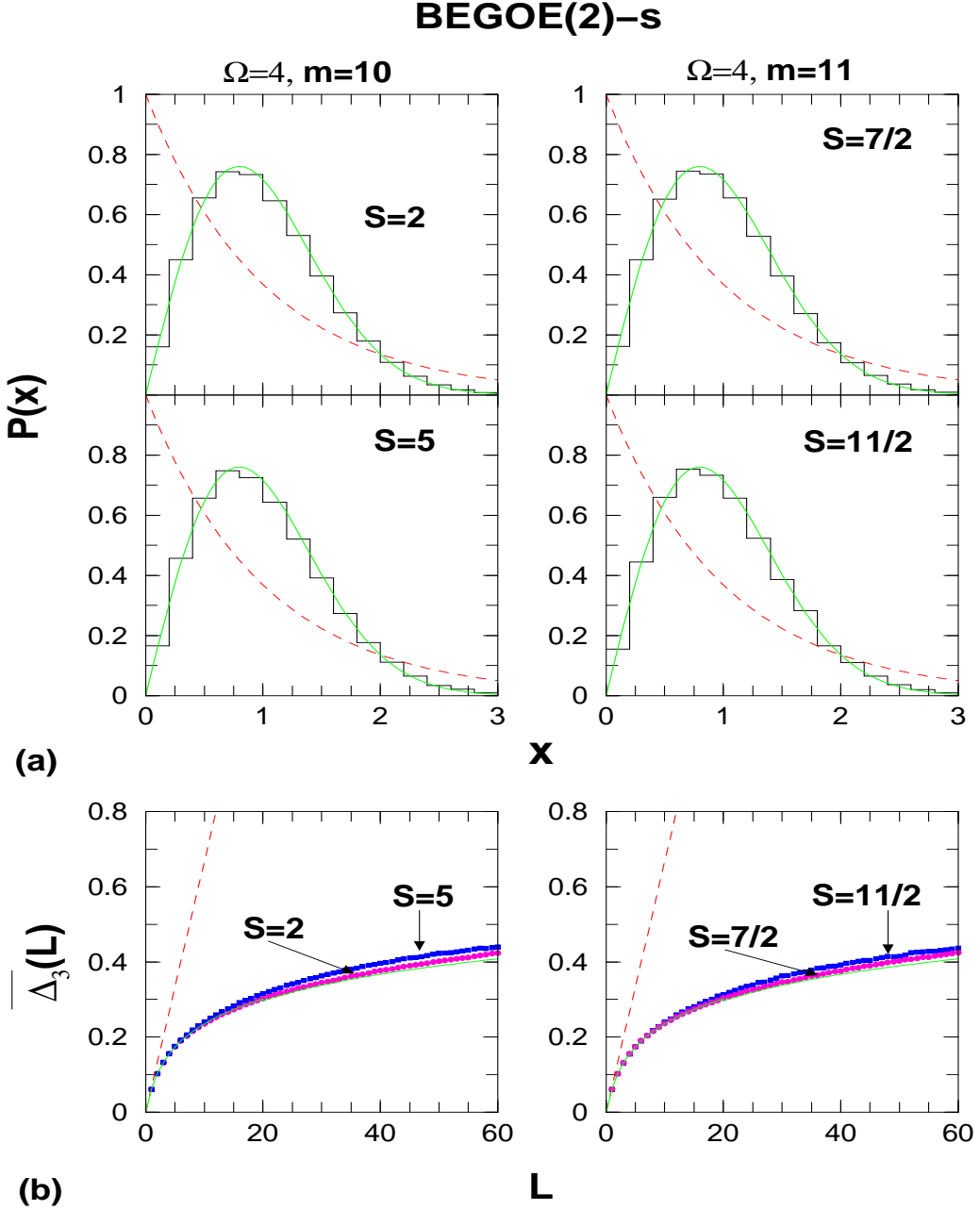}
\caption{(Color online) (a) Ensemble averaged nearest neighbor spacing 
distribution (NNSD) and (b) Dyson-Mehta statistic $\overline{\Delta_3}(L)$
vs  $L$ for $L \leq 60$. Results are for the same systems considered in
Fig. \ref{den}; first column gives the results for ($\Omega=4$, $m=10$) and
the second column for ($\Omega=4$, $m=11$) systems. The NNSD histograms from
BEGOE(2)-$\cs$ are  compared with Poisson (dashed red) and  GOE (Wigner)
forms (solid green) and similarly the $\overline{\Delta_3}(L)$ results.  In
the NNSD graphs, the bin-size is $0.2$ and  $x$ is the nearest neighbor
spacing in the units  of local mean spacing. See text and Fig. \ref{den} for
further details.}
\label{nnsd-del3}
\end{figure}

For the analysis of level fluctuations (equivalent to studying the two-point
function for the eigenvalues), each spectrum in the ensemble is unfolded
using  a sixth order polynomial correction to the Gaussian and then the 
smoothed density is $\overline{\eta(\widehat{E})}=\eta_{\cg}(\widehat{E}) \{
1+ \sum_{\zeta \geq 3}^{\zeta_0}  ( \zeta !)^{-1} S_{\zeta}
He_{\zeta}(\widehat{E})\}$ with $\zeta_0=6$ \cite{Le-08,PDPK}. The
parameters $S_{\zeta}$ are determined by minimizing
$\Delta^2=\sum_{i=1}^{d(\Omega,m,S)} [ F(E_i)-\overline{F(E)}]^2$. The
distribution function $F(E)=\int_{-\infty}^E \eta(x) dx$ and similarly
$\overline{F(E)}$ is defined. We require that the continuous function
$\overline{F(E)}$ passes through the mid-points of the jumps in the discrete
$F(E)$ and therefore, $F(E_i) = (i-1/2)$.  The ensemble averaged
$\Delta_{RMS}$ is $\sim 3$ for $\zeta_0=3$, $\sim 1$ for $\zeta_0=4$ and
$\sim 0.8$ for $\zeta_0=6$ with some variation with respect to $S$. As
$\Delta_{RMS}\sim 0.88$ for GOE, this implies GOE fluctuations set in when
we add 6th order corrections to the asymptotic Gaussian density. Using the
unfolded energy levels of all the members of the BEGOE(2)-$\cs$ ensemble,
the nearest neighbor spacing distribution  (NNSD) that gives information
about level repulsion and  the Dyson-Mehta $\overline{\Delta_3}(L)$
statistic that gives  information about spectral rigidity are studied.
Results for the same systems used in Fig. \ref{den} are shown in Fig.
\ref{nnsd-del3} for $S=2$ and $5$ (for other spins, the results are
similar).  In the calculations, middle 80\% of the eigenvalues from each
member are employed. It is clearly seen from the figures that the NNSD are
close to GOE (Wigner) form and the widths of the NNSD are $\sim 0.288$ (GOE
value is $\sim 0.272$). The $\overline{\Delta_3}(L)$ values show some
departures from GOE  for $L \gazz 30$ for $S=S_{max}$ and this could be
because the matrix dimensions are small for $S=S_{max}$ in our examples
(also the systems considered are not strictly in the dense limit and
numerical examples with  much larger $m$ and $\Omega$ with $m >> \Omega$ are
currently not feasible). It is useful to add that $S=S_{max}$ states are
important for boson systems with random interactions  as discussed in 
Sections IV-VI ahead. In conclusion, sixth order unfolding removes
essentially all the secular behavior and then the fluctuations follow 
closely  GOE. This is similar to the result known before for spinless 
boson systems \cite{Le-08,Ch-03}. 

\begin{figure}
\includegraphics[width=3in,height=6.5in]{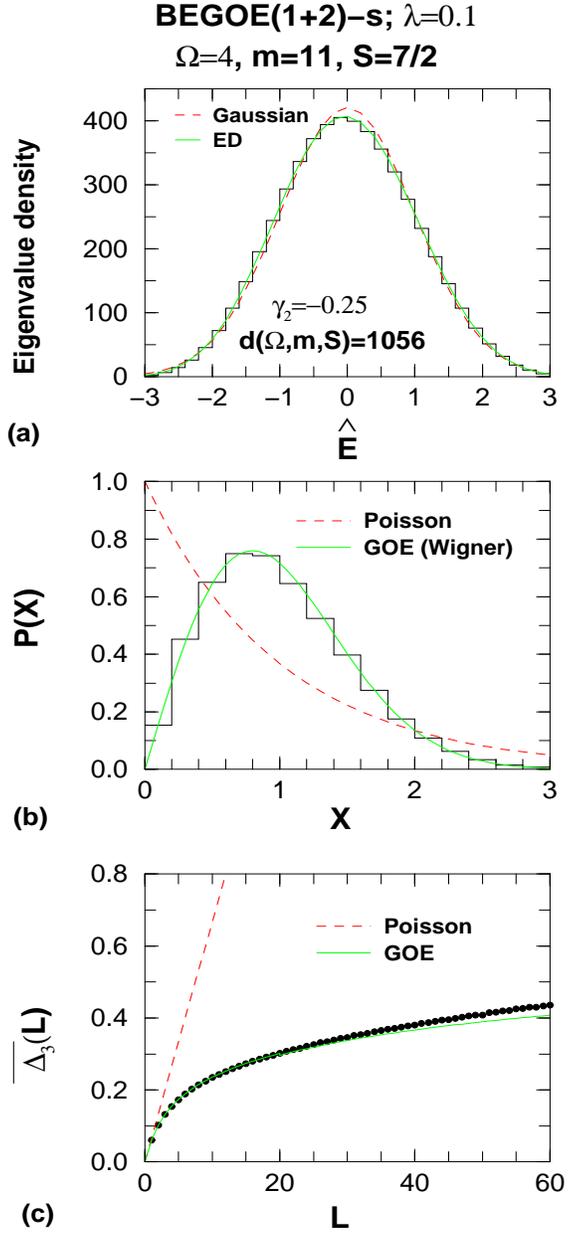}
\caption{(Color online) (a) Ensemble averaged eigenvalue density 
$\rho^{m,S}(\widehat{E})$, (b) NNSD and (c) 
$\overline{\Delta_3}(L)$ vs.  $L$ for a 100 member BEGOE(1+2)-$\cs$ 
ensemble for $\Omega=4$, $m=11$ and $S=7/2$ system with
$\lambda_0=\lambda_1=\lambda=0.1$ in Eq. (\ref{eq.begoe-s}). For all other
details, see text and Figs. \ref{den} and \ref{nnsd-del3}.}
\label{n4m11}
\end{figure}

\begin{figure}
\includegraphics[width=5.5in,height=6.5in]{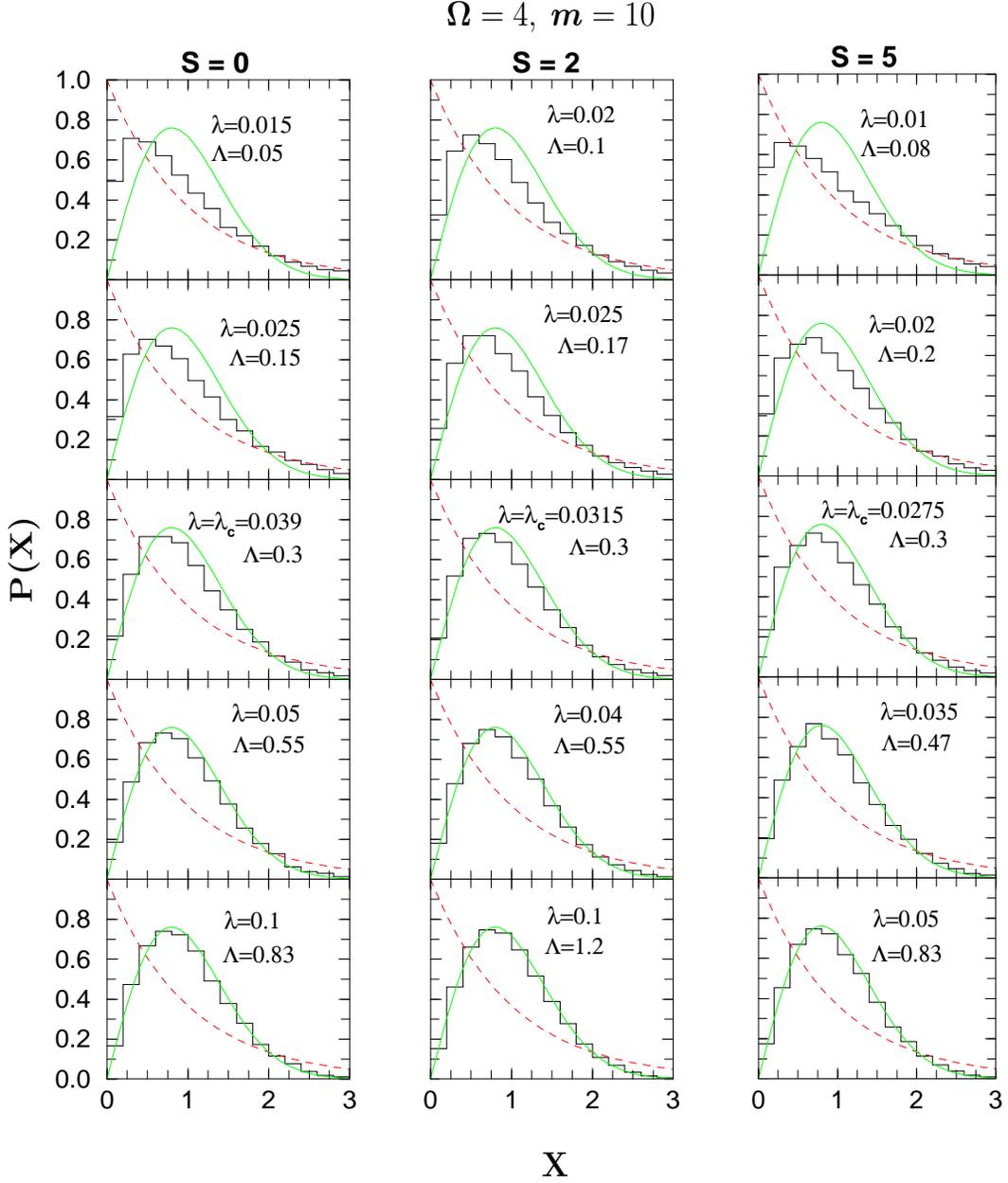}
\caption{(Color online) NNSD for a 100 member BEGOE(1+2)-$\cs$ ensemble 
with $\Omega=4$, $m=10$ and spins $S=0$, $2$ and $5$.  Calculated NNSD are
compared to the Poisson (red dashed) and Wigner (GOE) (green solid) forms. 
Values of the interaction
strength $\lambda$ and the transition parameter $\Lambda$ are given in  the
figure. The values of $\Lambda$ are deduced as discussed in \cite{Ma-10}. 
The chaos marker $\lambda_c$ corresponds to $\Lambda=0.3$ and its values, as
shown in the figure, are $0.039,\;0.0315,\;0.0275$ for $S=0,\;2$ and 5
respectively. Bin-size for the histograms is $0.2$.}
\label{ns-new}
\end{figure}

Going beyond BEGOE(2)-$\cs$, calculations are also carried out for
BEGOE(1+2)-$\cs$ systems using Eq. (\ref{eq.begoe-s}) with $\lambda_0 =
\lambda_1 = \lambda$. We have verified the Gaussian behavior for the
eigenvalue density for BEGOE(1+2)-$\cs$; an example is shown in  Fig.
\ref{n4m11}a. This result is essentially  independent of $\lambda$. In
addition, we have also verified that BEGOE(1+2)-$\cs$ also generates level
fluctuations close to GOE for $\lambda \gazz 0.1$ for $\Omega=4$ and $m=10$,
$11$ systems. Figure \ref{n4m11} shows an example with $\lambda=0.1$. Going
beyond this, in  Fig. \ref{ns-new}, we show the NNSD results, for a 100
member  BEGOE(1+2)-$\cs$ ensemble with $\Omega=4,\;m=10$ and total spins
$S=0,\;2$  and $5$, for $\lambda$ varying from 0.01 to 0.1 to demonstrate
that as $\lambda$ increases from zero,  there is generically Poisson to GOE
transition.  A similar study has been reported in \cite{Ma-10} for fermion
systems. As discussed there, for very small $\lambda$, the NNSD will be
Poisson (as we use sp energies to be $\epsilon_i=i+1/i$, the $\lambda=0$
limit will not give strictly a Poisson). Moreover, as discussed in detail in
\cite{Ma-10}, the variance of the NNSD  can be  written in terms of a
parameter $\Lambda$ ($\Lambda$ is a parameter in a 2$\times$2 random matrix
model that generates Poisson to GOE transition)  with $\Lambda=0$ giving
Poisson, $\Lambda \gazz 1$ GOE and $\Lambda=0.3$ the transition point
$\lambda_c$ that marks the onset of GOE fluctuations. We show in Fig.
\ref{ns-new}, for each $\lambda$, the deduced value of $\Lambda$ from the
variance of the NNSD  (Fig. \ref{nnsd-del3} gives results for $\lambda \to
\infty$ for $S = 2$ and 5).  As seen from the Fig. \ref{ns-new}, $\lambda_c=
0.039, \; 0.0315, \; 0.0275$ for $S=0$, $2$ and $5$ respectively. Thus
$\lambda_c$ decreases with increasing spin $S$ and this is opposite to the
situation for fermion systems. For a fixed $\Omega$ value, as discussed in
\cite{Ma-10}, the $\lambda_c$ is inversely proportional to $K$, where $K$ is
the number of many-particle states [defined by $h(1)$] that are directly
coupled by the two-body interaction. For fermion systems, $K$ is
proportional to the variance propagator but not for boson systems as
discussed in \cite{Ch-03}. At present, for BEGOE(1+2)-$\cs$ we don't have a
formula for $K$. However, if we use the variance propagator $Q(\Omega,m,S)$
for the boson systems [see Eq. (\ref{eq.varav1}) and Fig. \ref{prop} ahead],
then qualitatively we understand the decrease in $\lambda_c$ with increasing
spin.

Finally, it is well known that the Gaussian form for the eigenvalue density
is generic for embedded ensembles of spinless fermion \cite{Ko-01}  and
boson \cite{KP-80,Ch-03} systems. In addition,  ensemble averaged
fixed-($m,S$) eigenvalue densities for the fermion  EGOE(1+2)-$\cs$ are
shown to take  Gaussian form in \cite{Ko-06a,Ma-10}.  Hence, from the
results shown in Figs. \ref{den} and \ref{n4m11}a, it is  plausible  to
conclude that the Gaussian form is generic for EE (both bosonic and
fermionic) with good  quantum numbers. With the eigenvalue density being
close to Gaussian, it is useful to derive formulas for the energy centroids
and ensemble averaged spectral variances. These in turn, as discussed ahead,
will also allow us to study the lowest two moments of the two-point
function. From now on, we will drop the \lq{hat}\rq  over the operators $H$,
$h(1)$ and $V(2)$.

\section{Energy centroids, spectral variances and ensemble averaged spectral
variances and covariances}

\subsection{Propagation formulas for energy centroids and
spectral variances}

Given a general (1+2)-body Hamiltonian $H=h(1)+V(2)$, which is a typical
member of BEGOE(1+2)-$\cs$, the energy centroids will be polynomials in  the
number operator and the $S^2$ operator. As $H$ is of maximum body rank 2,
the polynomial form for the energy centroids is $\lan H \ran^{m,S} =
E_c(m,S) = a_0 +a_1 m +a_2 m^2 + a_3 S(S+1)$. Solving for the $a$'s in terms
of the centroids in one and two particle spaces, the propagation formula for
the energy  centroids is,
\be
\barr{rcl}
\lan H \ran^{m,S} = E_c(m,S) & = & \l[\lan h(1) \ran^{1,\spin}\r]\;m +
\lambda_0\,
\lan\lan V^{s=0}(2) \ran\ran^{2,0}\;\dis\frac{P^0(m,S)}{4\Omega 
(\Omega-1)} \\
& + & \lambda_1\,\lan\lan V^{s=1}(2) \ran\ran^{2,1}\;\dis\frac{P^1(m,S)}{
4\Omega (\Omega+1)}\;\;;
\\
P^0(m,S) & = & \l[ m(m+2) - 4S(S+1)\r]\;\;,\\
P^1(m,S) & = & \l[3m(m-2) + 4S(S+1)\r]\;\;, \\
\lan h(1) \ran^{1,\spin} & =
&\overline{\epsilon}=\Omega^{-1}\;\dis\sum_{i=1}^{\Omega}\;
\epsilon_i\;\;, \\
\lan\lan V^{s=0}(2) \ran\ran^{2,0} & = & \dis\sum_{i
< j}\;V^{s=0}_{ijij}\;,\;\;\;
\lan\lan V^{s=1}(2) \ran\ran^{2,1} = \dis\sum_{i \leq j}\;V^{s=1}_{ijij}\;.
\earr \label{eq.bcent}
\ee
For the energy centroid of a two-body Hamiltonian [member of a
BEGOE(2)-$\cs$], the $h(1)$ part in Eq. (\ref{eq.bcent}) will be absent.

Just as for the energy centroids, polynomial form for the spectral variances
$$
\sigma_{H=h(1)+V(2)}^2(m,S) = \lan H^2 \ran^{m,S} - \l[E_c(m,S)\r]^2
$$ 
is $\sum_{p=0}^4 a_p m^p + \sum_{q=0}^2 b_q m^q S(S+1) + c_0 [S(S+1)]^2$. It
is well known that the propagation formulas for fermion systems will give
the formulas for the corresponding boson systems by applying $\Omega
\rightarrow -\Omega$ transformation \cite{Ko-79,KP-80,Ko-81,Cv-82,Ko-05}.
Applying this transformation to the propagation equation for the spectral
variances for fermion systems with spin given by Eq. (8) of \cite{Ko-06a},
we obtain the propagation equation for $\sigma_{H=h(1)+V(2)}^2(m,S)$ in
terms of inputs that contain the single particle energies $\epsilon_i$
defining $h(1)$ and the two particle matrix elements $V_{ijkl}^s$. The final
result is,
\be
\barr{l}
\sigma_{H=h(1)+V(2)}^2(m,S) = \lan H^2 \ran^{m,S} - \l[E_c(m,S)\r]^2 \\
\\
= \dis\frac{(\Omega-2)mm^\star+2\Omega \;\lan S^2 \ran}{ (\Omega-1) \Omega
(\Omega+1)}\;\;\dis\sum_i\;{\tilde{\epsilon}}_i^2 \\
+ \dis\frac{m^\star P^0(m,S)}{2  (\Omega-1) \Omega
(\Omega+1)}\;\;\dis\sum_i\;{\tilde{\epsilon}}_i \lambda_{i,i}(0) \\
+ \dis\frac{(\Omega-2)m^\star P^1(m,S) +
8 \Omega (m-1) \lan S^2 \ran}{2 (\Omega-1) \Omega (\Omega+1) (\Omega+2)}\;\;
\dis\sum_i\;{\tilde{\epsilon}}_i \lambda_{i,i}(1) \\
+ P^{\nu=1,s=0}(m,S)\;\; \dis\sum_{i,j}\; \lambda_{i,j}^2(0)
+ P^{\nu=1,s=1}(m,S)\;\; \dis\sum_{i,j}\;\lambda_{i,j}^2(1) \\
+ \dis\frac{P^2(m,S) P^0(m,S)}
{4  (\Omega-1) \Omega (\Omega+1) (\Omega+2)}\;\; \dis\sum_{i,j}\;
\lambda_{i,j}(0) \lambda_{i,j}(1) \\
+ P^{\nu=2,s=0}(m,S)\,\lan \l(V^{\nu=2,s=0}\r)^2 \ran^{2,0} +
P^{\nu=2,s=1}(m,S)\,\lan \l(V^{\nu=2,s=1}\r)^2 \ran^{2,1}\;;
\earr \label{eq.bvar}
\ee
The propagators $P^{\nu,s}$'s, which are used later, are
\be
\barr{l}
P^{\nu=1,s=0}(m,S) = \dis\frac{\l[(m+2)m^\star/2 - \lan S^2 \ran\r]
P^0(m,S)}
{8 (\Omega-2) (\Omega-1) \Omega (\Omega+1) } \;,\\
P^{\nu=1,s=1}(m,S) =\dis\frac{8\Omega(m-1)(\Omega+2m-4) \lan S^2 \ran +
(\Omega-2) P^2(m,S) P^1(m,S)}
{8  (\Omega-1) \Omega (\Omega+1) (\Omega+2)^2} \;,\\
P^{\nu=2,s=0}(m,S) = \l[m^\star(m^\star-1) - \lan S^2 \ran\r]
P^0(m,S)/[8 \Omega (\Omega+1)]\;,\\\\
P^{\nu=2,s=1}(m,S) = \l\{\l[\lan S^2 \ran\r]^2 (3\Omega^2+7\Omega+6)/2 +
3m(m-2) m^\star (m^\star+1) \times \r. \\
\l. (\Omega-1)(\Omega-2)/8 + \l[\lan S^2 \ran/2\r]\l[(5\Omega+3)
(\Omega-2)m m^\star +
\Omega(\Omega-1)(\Omega+1)(\Omega-6)\r]\r.\}/ \\
\l[(\Omega-1) \Omega (\Omega+2)(\Omega+3)\r]\;; \\
P^2(m,S) = 3(m-2) m^\star/2 + \lan S^2 \ran\;,\;\;\;\;m^\star =
\Omega+m/2\;,\;\;\;\; \lan S^2 \ran =S(S+1)\;.
\earr \label{eq.bvar1}
\ee
The inputs in Eq. (\ref{eq.bvar}) are given by,
\be
\barr{l}
{\tilde{\epsilon}}_i=\epsilon_i -\overline{\epsilon} \;,\\
\lambda_{i,i}(s) =  \dis\sum_j\;V_{ijij}^s\;(1+\delta_{ij})
\;-\;(\Omega)^{-1} \;\dis\sum_{k,l}\;V^s_{klkl}\;(1+\delta_{kl}) \;, \\
\lambda_{i,j}(s) = \dis\sum_k\;\dis\sqrt{(1+\delta_{ki})(1+\delta_{kj})}\,
V^s_{kikj}\;\;\;\mbox{for}\;\;\;i \neq j \;,\\
V^{\nu=2,s}_{ijij} =  V^s_{ijij} - \l[\lan V(2)\ran^{2,s} +
(\lambda_{i,i}(s) + \lambda_{j,j}(s))\l(\Omega-2(-1)^s\r)^{-1}\r] \;,\\
V^{\nu=2,s}_{kikj} = V^s_{kikj} - \l(\Omega-2(-1)^s\r)^{-1}\,
\dis\sqrt{(1+\delta_{ki})(1+\delta_{kj})}\,\lambda^s_{i,j}
\;\;\;\mbox{for}\;\;\;i \neq j \;,\\
V^{\nu=2,s}_{ijkl} = V^s_{ijkl}\;\;\;\mbox{for all other cases}\;.
\earr \label{eq.bvar2}
\ee
Eqs. (\ref{eq.bcent}) and (\ref{eq.bvar}) can be applied to individual
members of the BEGOE(1+2) ensemble. On the other hand, it is possible to use
these to obtain ensemble averaged spectral variances and ensemble averaged
covariances in energy centroids just as it was done before for fermion
systems \cite{Ma-10}. Now we  will consider these.

\subsection{Ensemble averaged spectral variances for BEGOE(2)-$\cs$}
\label{spvar}

In this subsection, we restrict to $H=V(2)$, i.e. BEGOE(2)-$\cs$ and
consider BEGOE(1+2)-$\cs$ at the end.

For the ensemble averaged spectral variances generated by $H$, only the
fourth, fifth, seventh and eighth terms in Eq. (\ref{eq.bvar}) will
contribute. Evaluating the ensemble averages of the inputs in these four
terms, we obtain,
\be
\barr{l}
\overline{\dis\sum_{i,j}\;\lambda_{i,j}^2(0)} =
\lambda_0^2 (\Omega-1)(\Omega-2)(\Omega+2) \;,\\
\overline{\dis\sum_{i,j}\;\lambda_{i,j}^2(1)} =
\lambda_1^2 (\Omega-1)(\Omega+2)^2 \;,\\
\overline{\lan \l(H^{\nu=2,s=0}\r)^2 \ran^{2,0}} =
\lambda_0^2 \dis\frac{(\Omega-3)(\Omega^2+\Omega+2)}{2(\Omega-1)} \;,\\
\overline{\lan \l(H^{\nu=2,s=1}\r)^2 \ran^{2,1}} =
\lambda_1^2 \dis\frac{(\Omega-1)(\Omega+2)}{2}\;.
\earr \label{eq.varinp}
\ee
Note that these inputs follow from the results for EGOE(2)-$\cs$ for
fermions given in \cite{Ma-10} by interchanging $s=0$ with $s=1$.
Now the final expression for the ensemble averaged variances is
\be
\barr{rcl}
\overline{\sigma^2_{H}(m,S)} & = & \dis\sum_{s=0,1}
\lambda_s^2 (\Omega-1)(\Omega-(-1)^s 2)(\Omega+2)\;P^{\nu=1,s}(m,S) \\
& + & \lambda_0^2 \dis\frac{(\Omega-3)(\Omega^2+\Omega+2)}{2(\Omega-1)}
\; P^{\nu=2,s=0}(m,S) \\
& + & \lambda_1^2 \dis\frac{(\Omega-1)(\Omega+2)}{2}\; P^{\nu=2,s=1}(m,S)\;.
\earr \label{eq.varav}
\ee
In most of the numerical calculations, we employ $\lambda_0 = \lambda_1
= \lambda$ and then $\overline{\sigma^2_{H}(m,S)}$ takes the form,
\be
\overline{\sigma^2_{H}(m,S)} \stackrel{\lambda_0 = \lambda_1 =
\lambda}{\longrightarrow} \lambda^2 Q(\Omega,m,S)\;.
\label{eq.varav1}
\ee
Expression for the variance propagator $Q(\Omega,m,S)$ follows easily from
Eqs. (\ref{eq.bcent}), (\ref{eq.bvar1}) and (\ref{eq.varav}). In Fig.
\ref{prop}, we show a plot of $Q(\Omega,m,S)/Q(\Omega,m,S_{max})$ vs
$S/S_{max}$ for various $\Omega$ and $m$ values. It is clearly seen that the
propagator value increases as spin increases and this is just opposite to
the result for fermion systems \cite{Ma-10}. An important consequence of
this is BEGOE(2)-$\cs$ gives ground states with $S=S_{max}$ [for fermion
EGOE(2)-$\cs$, the ground states with random interactions have $S=0$]. This
result follows from the Jacquod and Stone \cite{JS-01} criterion and
according to this (from the assumption of Gaussian form for the eigenvalue
densities), the gs energy $E_{gs}$ is given by $E_{gs}
\propto -\dis\sqrt{ \overline{\sigma^2_{H}(m,S)}}$.

Before proceeding further, let us remark that for the BEGOE(1+2)-$\cs$
Hamiltonian $\{H\}=h(1)+\{V(2)\}$, assuming that $h(1)$ is fixed, we have
$\overline{\sigma^2_{H}} = \sigma^2_{h(1)} + \overline{\sigma^2_{V(2)}}$.
The first term $\sigma^2_{h(1)}$ is given by the first term of Eq.
(\ref{eq.bvar}) and the second term is given by Eq. (\ref{eq.varav}).
In the situation  $h(1)$ is represented by an ensemble independent of 
$\{V(2)\}$, we have to replace $\sigma^2_{h(1)}$ by
$\overline{\sigma^2_{h(1)}}$ in $\overline{\sigma^2_H}$.

\subsection{Ensemble averaged covariances in energy centroids and spectral
variances for BEGOE(2)-$\cs$}

Normalized covariances in energy centroids and spectral variances are
defined by
\be
\barr{rcl}
\Sigma_{11}(m,S:m^\pr,S^\pr) & = &
\dis\frac{\overline{\lan H \ran^{m,S}\lan H
\ran^{m^\pr,S^\pr}} - \l\{\overline{\lan H \ran^{m,S}}\r\} \;\;
\l\{\overline{\lan H \ran^{m^\pr,S^\pr}} \r\}
}{\dis\sqrt{\l\{\overline{\lan H^2 \ran^{m,S}}\r\}\;\;
\l\{\overline{\lan H^2 \ran^{m^\pr,S^\pr}}\r\}}}\;,\\
\\
\Sigma_{22}(m,S:m^\pr,S^\pr) & = &
\dis\frac{\overline{\lan H^2 \ran^{m,S}\lan H^2
\ran^{m^\pr,S^\pr}} - \l\{\overline{\lan H^2 \ran^{m,S}}\r\} \;\;
\l\{\overline{\lan H^2 \ran^{m^\pr,S^\pr}}\r\}
}{\l\{\overline{\lan H^2 \ran^{m,S}}\r\}\;\;
\l\{\overline{\lan H^2 \ran^{m^\pr,S^\pr}}\r\}}\;.
\earr \label{eq.covec}
\ee
These define the lowest two moments of the two-point function,
\be
\cS^{\Omega,m,S:\Omega,m^\pr,S^\pr}(E,W) = \overline{\rho^{m,S}(E)
\rho^{m^\pr,S^\pr}(W)} -
\l\{\overline{\rho^{m,S}(E)}\r\}\;\;
\l\{\overline{\rho^{m^\pr,S^\pr}(W)}\r\}\;.
\label{eq.twopt}
\ee
For  $(m,S) = (m^\pr,S^\pr)$ they will give information about fluctuations
and in particular about level motion in the ensemble \cite{PDPK}. For $(m,S)
\neq (m^\pr,S^\pr)$, the covariances (cross correlations) are non-zero for
BEGOE while they will be zero for independent GOE representation for the $m$
boson Hamiltonian matrices with different $m$ or $S$. Note that the 
$\Omega$ value has to be  same for both $(m,S)$ and $(m^\pr,S^\pr)$ systems
so that the Hamiltonian in two-particle spaces remains same. Now we will
discuss analytical  and numerical results  for $\Sigma_{11}$ and numerical
results for $\Sigma_{22}$ for large  values of $(\Omega,m)$ and they are
obtained using the results in Section IV.

\begin{figure}
\includegraphics[width=6in]{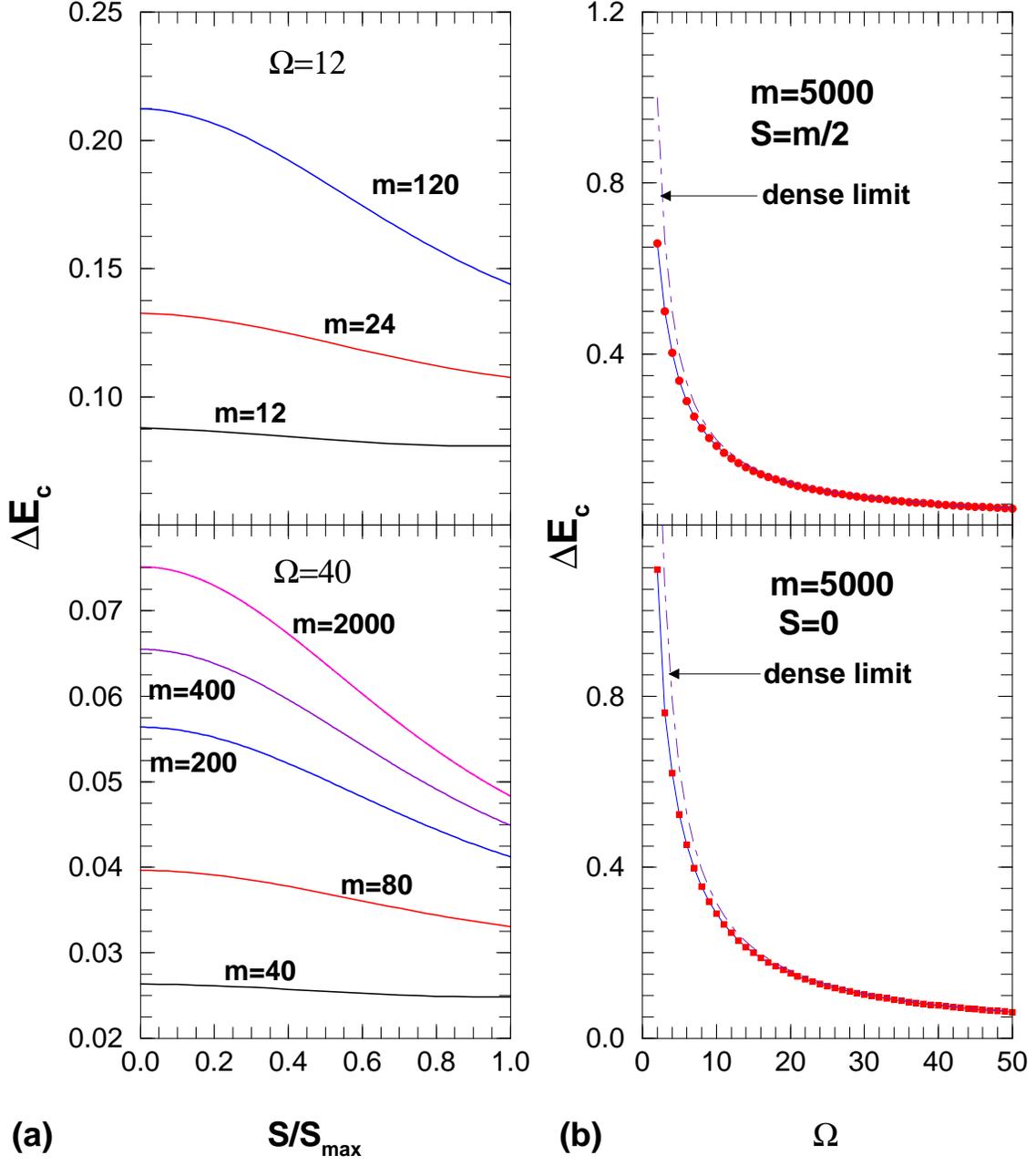}
\caption{(Color online) (a) Self-correlations $\Sigma^{1/2}_{11}$ in energy
centroids, giving  width $\Delta E_c$ of the fluctuations in energy
centroids scaled to the spectrum width, as a function of spin $S$ for
different values of $m$  and $\Omega$.  (b) Self correlations as a function
of $\Omega$ for 5000 bosons  with minimum spin $(S=0)$ and maximum spin
$(S=2500)$. Dense limit (dot-dashed) curves for 
$S=0$ and $S=m/2$ in (b) correspond to the results given by 
Eq. (\ref{asymp4}). See text for details.}
\label{sig11-s}
\end{figure}

\begin{figure}
\includegraphics[width=6in]{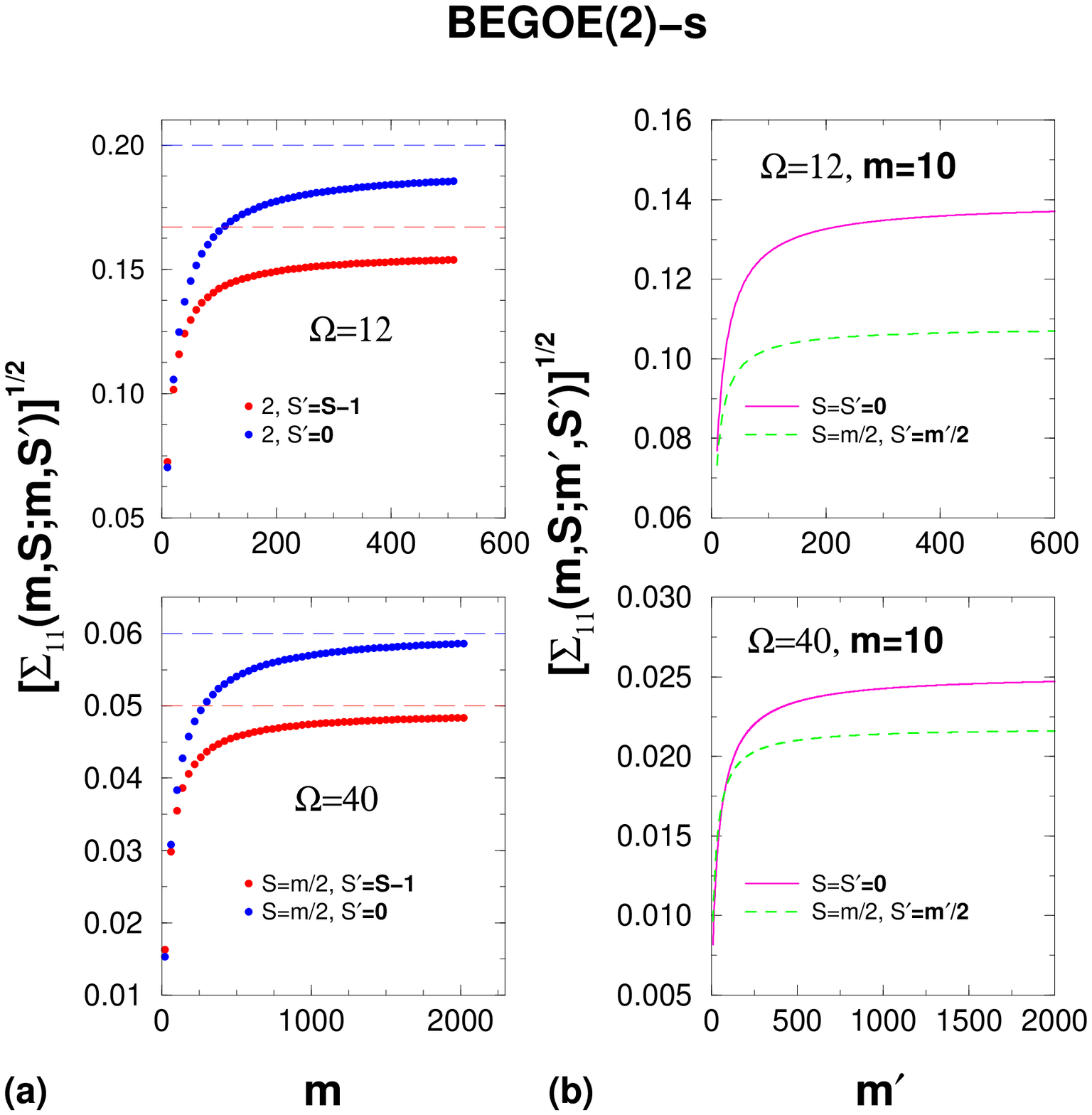}
\caption{(Color online) Cross-correlations $\Sigma^{1/2}_{11}$ in energy
centroids for  various BEGOE(2)-$\cs$  systems. (a) $\Sigma^{1/2}_{11}$ vs
$m$ with $m=m^\pr$ but different spins  $(S \neq S^\pr)$. (b)
$\Sigma^{1/2}_{11}$ vs $m^\pr$ with $m=10$ and  $S=S^\pr=0$ and
$S=5,S^\pr=m^\pr/2$. The dashed lines in (a) are the dense limit results.
See text for details.}
\label{sig11-c}
\end{figure}

Trivially, the ensemble average of the energy centroids $E_c(m,S)$ will be
zero [note that $H$ is two-body for BEGOE(2)-$\cs$], i.e. $\overline{\lan
H \ran^{m,S}}=0$. However the covariances in the energy centroids of $H$ are
non-zero and Eq. (\ref{eq.bcent}) gives,
\be
\barr{l}
\overline{\lan H \ran^{m,S}\lan H \ran^{m^\pr,S^\pr}} = \\
\dis\frac{\lambda^2_0}{16\Omega(\Omega-1)}P^0(m,S)\,P^0(m^\pr,S^\pr) +
\dis\frac{\lambda^2_1}{16\Omega(\Omega+1)}P^1(m,S)\,P^1(m^\pr,S^\pr)\;.
\earr \label{eq.eccor}
\ee
Equations (\ref{eq.varav}), (\ref{eq.varav1}) and (\ref{eq.eccor}) allow us
to calculate $\Sigma_{11}$ for any $(\Omega,m,S)$. For $m=m^\pr$ and
$S=S^\pr$, the $[\Sigma_{11}]^{1/2}$ gives the width $\Delta E_c$ of the
fluctuations in the energy centroids. In the numerical calculations, we use
$\lambda_0=\lambda_1=\lambda$ and therefore, $\Sigma_{11}$ and $\Sigma_{22}$
are independent of $\lambda$. Figure \ref{sig11-s} gives some numerical
results for $\Delta E_c$ and it is seen that : (i) for $m >> \Omega$, the
${\Delta}E_c$ is $\sim 20$\% for $S=0$ and it goes down to $\sim 15$\% for
$S=S_{max}=m/2$ for $\Omega=12$; (ii) going from $\Omega=12$ to $40$,
$\Delta E_c$ decreases to $\sim 2-7$\%; (iii) for fixed $(m,\Omega)$, there
is decrease in $\Delta E_c$ with increasing $S$ value; (iv) for fixed
$(m,S)$ and very large $m$ value, there is a sharp decrease in $\Delta E_c$
with increasing $\Omega$ up to $\Omega \sim 20$ and then it slowly converges
to zero. It is possible to understand these results and the results for
cross correlations $[\Sigma_{11}(m,S:m^\pr,S^\pr)]^{1/2}$, with $(m,S) \neq
(m^\pr,S^\pr)$ as shown in Fig. \ref{sig11-c}, using the asymptotic
structure of $Q(\Omega,m,S)$.

\begin{figure}
\includegraphics[width=6in]{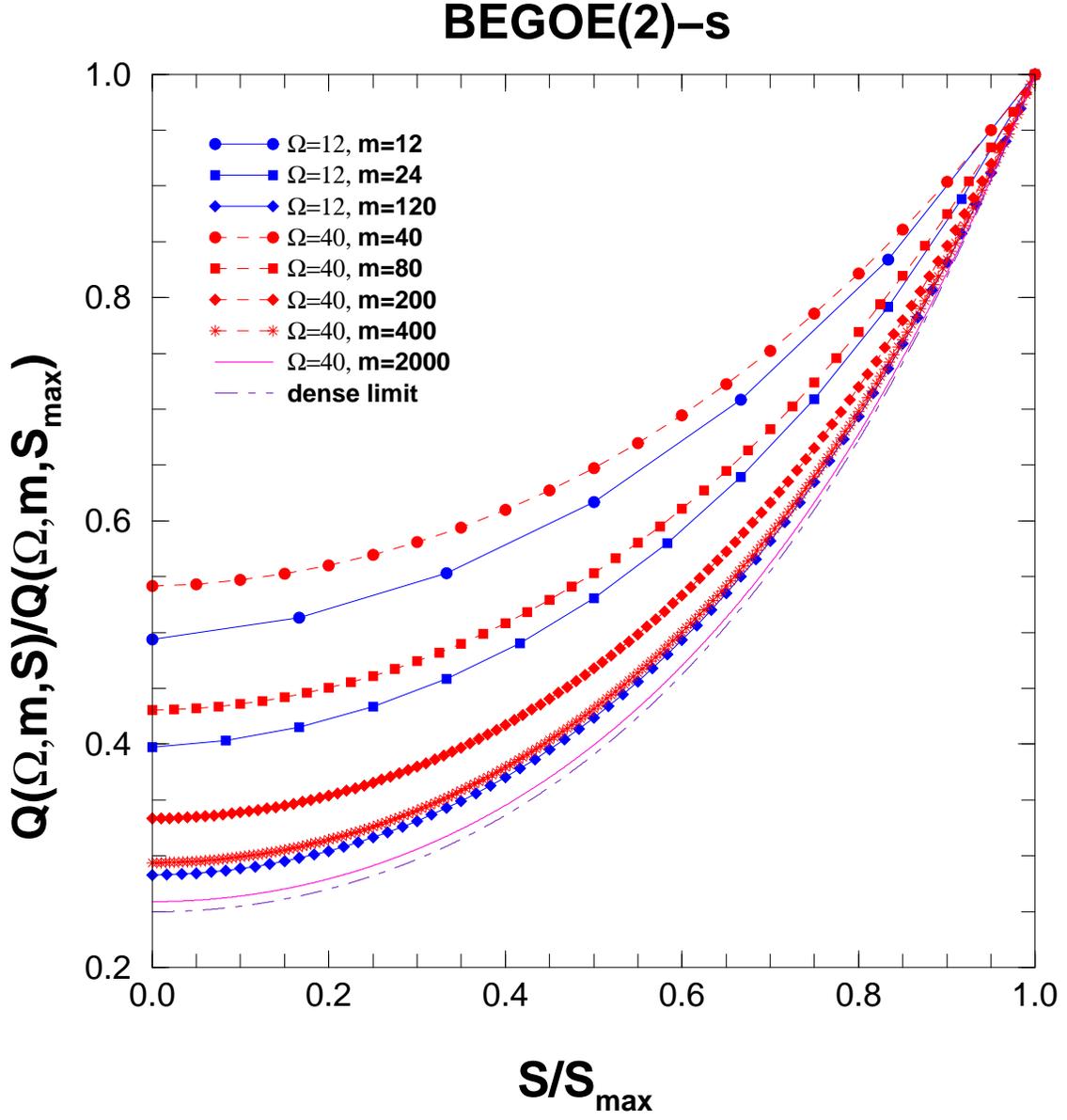}
\caption{(Color online) BEGOE(2)-$\cs$ variance propagator 
Q($\Omega$,$m$,$S$)/Q($\Omega$, $m$,$S_{max}$) vs $S/S_{max}$ for various
values of $\Omega$ and $m$.   Formula for Q($\Omega$,$m$,$S$) follows from
Eqs. (\ref{eq.bvar1}), (\ref{eq.varav}) and  (\ref{eq.varav1}). Note that
the results in the figure are for $\lambda_0=\lambda_1=\lambda$ in Eq.
(\ref{eq.v2}) and therefore independent of $\lambda$. Dense limit
(dot-dashed) curve corresponds to the result due to Eq. (\ref{asymp2}) 
with $m=2000$.}
\label{prop}
\end{figure}

Let us consider the dense limit defined by $m \rightarrow \infty$, $\Omega
\rightarrow \infty$ and $m/\Omega \rightarrow \infty$.
Firstly the $P^{{\nu},s}(m,S)$ in Eq. (\ref{eq.bvar1}) take the simpler
forms, with $S^2=S(S+1)$,
\be
\barr{l}
P^{\nu=1,s=0} = \dis\frac{\l(m^2 - 4S^2\r)^2}{32 \Omega^4}\;,\;\;\;\;
P^{\nu=1,s=1} = \dis\frac{64m^2S^2 \l(3m^2 + 4S^2\r)^2}{32 \Omega^4}\;,\\
\\
P^{\nu=2,s=0} = \dis\frac{\l(m^2 - 4S^2\r)^2}{32 \Omega^2}\;,\;\;\;\;
P^{\nu=2,s=1} = \dis\frac{3m^4 + 40 m^2S^2 + 48(S^2)^2}{32 \Omega^2}\;.
\earr \label{asymp1}
\ee
Using these in Eq. (\ref{eq.varav}), with $\lambda_0=\lambda_1=\lambda$, we
have
\be
\barr{l}
\overline{\sigma_H^2(m,S)} = \lambda^2 \dis\frac{\l(m^2 +
4S^2\r)^2}{16} \;;\\
\Rightarrow \overline{\sigma_H^2(m,S)} / \overline{\sigma_H^2(m,S_{max})} =
\l[ \dis\frac{m/(m+2)+S^2/S_{max}^2}{m/(m+2)+1} \r]^2 \;.
\earr \label{asymp2}
\ee
The dense limit result given by Eq. (\ref{asymp2}) with $m=2000$  is
compared with the exact results in Fig. \ref{prop}. Firstly, it should be
noted that for the applicability of Eq. (\ref{asymp2}), $\Omega$ should be
sufficiently large and $m >> \Omega$. Also, the result is independent of
$\Omega$. Comparing with the $\Omega = 12$ and $\Omega = 40$ results, it is
seen that the dense limit result is very close to the $\Omega = 40$ results
for $m \gazz 200$. Thus for sufficiently large value of $\Omega$ and $m
\gazz 5\Omega$, the dense limit result describes quite well the exact 
results.

Simplifying $\overline{\lan H \ran^{m,S} \lan H \ran^{m^\pr,S^\pr}}$ gives 
in the dilute limit,
\be
\barr{l}
\overline{\lan H \ran^{m,S}\lan H\ran^{m^\pr,S^\pr}} = \\
\dis\frac{\lambda^2}{16 \Omega^2}\l[\l(m^2 -4S^2\r)\l\{(m^\pr)^2
-4(S^\pr)^2\r\} + \l(3m^2 +4S^2\r)\l\{3(m^\pr)^2
+4(S^\pr)^2\r\}\r]\;.
\earr \label{asymp3}
\ee
Then $[\Sigma_{11}]^{1/2}$, with $m=m^\pr$ and $S=S^\pr$ (for
$\lambda_0=\lambda_1$) giving $\Delta E_c$, is
\be
[\Sigma_{11}]^{1/2} = \Delta E_c = \dis\frac{\dis\sqrt{2(5m^4 +8m^2S^2 +
16(S^2)^2)}}{\Omega\l(m^2+4S^2\r)}\;.
\label{asymp4}
\ee
Eq. (\ref{asymp4}) gives $[\Sigma_{11}]^{1/2}$ to be $\sqrt{10}/\Omega$ and
$2/\Omega$ for $S=0$ and $S=S_{max}$ and these dense limit results are well
verified by the results in Fig. \ref{sig11-s}b.  Similarly, Eqs.
(\ref{asymp2}) and (\ref{asymp3}) will give $[\Sigma_{11}]^{1/2}$ to be
$\sqrt{6}/\Omega$ for $(m=m^\pr : S=S_{max},S^\pr=0)$ and $2/\Omega$ for
$(m=m^\pr : S=S_{max},S^\pr=S_{max}-1)$. The upper and lower dashed lines in
Fig. \ref{sig11-c}a  for $\Omega = 12$ (similarly for $\Omega = 40$)
correspond to these two dense limit results respectively. It is seen that
the dense limit results are close to exact results for $\Omega  = 40$ but
there are deviations for $\Omega = 12$. Also, for $\Omega = 40$, the
agreements are good only for $m \gazz 80$ and these are similar to the
results discussed earlier with reference to Fig. \ref{prop}.

\begin{figure}
\includegraphics[width=6in]{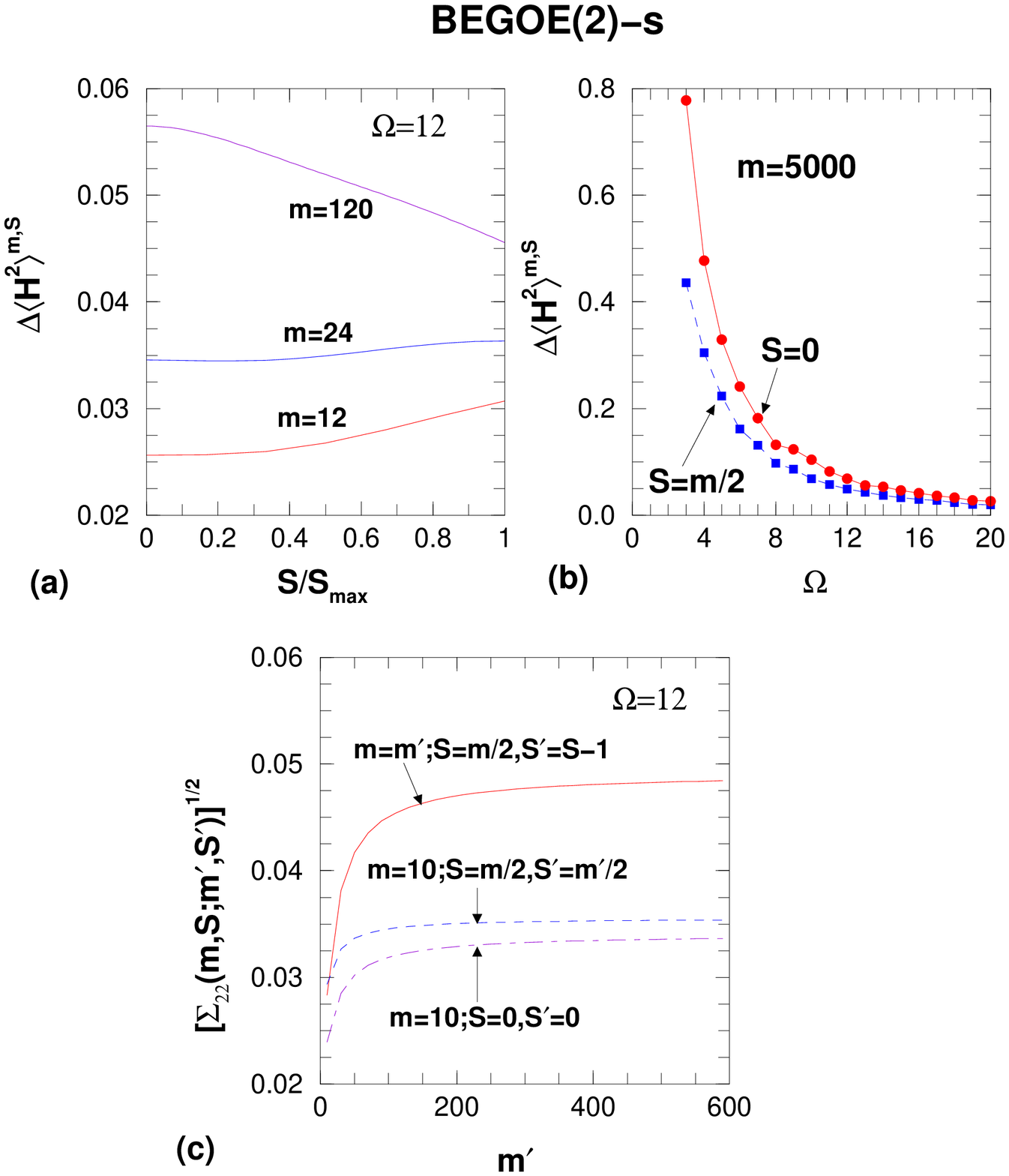}
\caption{(Color online) Correlations in spectral variances
$\Sigma^{1/2}_{22}$ for various  BEGOE(2)-$\cs$ systems. (a)
Self-correlations, giving width $\Delta\lan H^2 \ran^{m,S}$  of the 
spectral variances,  as a function of spin $S$ for $m=12$, $24$ and $120$
with $\Omega=12$. (b) Self-correlations as a function of $\Omega$ for 5000
bosons with  $S=0$ and $2500$. (c) Three examples for cross-correlation in 
spectral variances with same or different particle numbers  and same or
different spins. All the results are obtained using 500 member  ensembles. 
See text for details.}
\label{sig22}
\end{figure}

Unlike for the covariances in energy centroids, we do not have at present
complete analytical formulation for the covariances in spectral variances.
However, for a given member of BEGOE(2)-$\cs$, generating numerically (on a
computer) the ensembles $\{V^{s=0}(2)\}$ and $\{V^{s=1}(2)\}$ and applying
Eqs. (\ref{eq.bcent}) and (\ref{eq.bvar}) to each member of the ensemble 
will give $\overline{\lan H^2\ran^{m.S}} =  \overline{\sigma^2(m,S)} +
\overline{\l[E_c(m,S)\r]^2}$. This procedure has been used with 500 members
and results for $\Sigma_{22}$ are obtained for various $({\Omega},m,S)$
values. For some examples, results are shown in Fig. \ref{sig22} for both
self correlations giving the width $\Delta \lan H^2 \ran^{m,S}$ of variances
and cross correlations $[\Sigma_{22}]^{1/2}$ with $(m,S) \neq
(m^\pr,S^\pr)$. It is seen that $[\Sigma_{22}]^{1/2}$ are always much
smaller than $[\Sigma_{11}]^{1/2}$ just as for EGOE(2) for spinless fermion
systems \cite{Ko-06a}. It is seen from Fig. \ref{sig22}a that for
$\Omega=12$, width of the fluctuations in the variances $\lan H^2\ran^{m,S}$
are $\sim 3-5${\%}. Similarly for large $m$, with $\Omega$ very small, the
widths are quite large but they decrease fast with increasing $\Omega$ as
seen from Fig. \ref{sig22}b. Finally, for $\Omega=12$, the cross
correlations are $\sim 4$\%. 

Besides the moments $\Sigma_{11}$ and $\Sigma_{22}$,  it is possible to
numerically construct the two-point function
$\cS^{\Omega,m,S:\Omega,m^\pr,S^\pr}(E,W)$ using the eigenvalues from the
BEGOE(2)-$\cs$ Hamiltonian matrices in small examples. We have carried out
the calculations for $\cS^{4,10,0:4,10,1}(E,W)$ using a 500 member
BEGOE(2)-$\cs$ ensemble. It is seen that the structure of $\cS(E,W)$ is
similar to the nuclear shell model examples reported in \cite{PW-06} and
EGOE(1+2)-$\cs$ example in \cite{Ma-10c}. The maximum value of 
$\cS^{4,10,0:4,10,1}(E,W)$ is found to be $\sim 7$\% of
$\overline{\rho^{10,0}(E)}\,\times\,\overline{\rho^{10,1}(W)}$. Let us add
that it is important to identify measures involving $\Sigma_{11}$ and
$\Sigma_{22}$ and also $\cS^{\Omega,m,S:\Omega,m^\pr,S^\pr}(E,W)$, $(m,S)
\neq (m^\pr,S^\pr)$ that can be tested using some experiments so that
evidence for BEGOE(2) operation in real quantum systems can be
established.  

\section{Preponderance of $S_{max}=m/2$ ground states and natural spin order
: role of exchange interaction}

\subsection{Introduction to regular structures with random interactions}

Johnson et al \cite{Jo-98} discovered in 1998 that the  nuclear shell model
with random interactions generates, with high probability, $0^+$ ground
states in even-even nuclei (also generates odd-even staggering in binding
energies, the seniority pairing gap etc.) and similarly, Bijker and Frank
\cite{BF-00} found that the interacting boson model ($sd$IBM) of atomic
nuclei [in this model, one considers identical bosons carrying angular
momentum $\ell=0$ (called $s$ bosons) and $\ell=2$ (called $d$ bosons)] with
random interactions generates vibrational and rotational structures with
high probability. Starting with these, there are now many studies on regular
structures in many-body systems generated by random interactions. See for
example \cite{Zh-04,Zel-04,We-09} for reviews on the subject. More
recently,  the effect of random interactions in the $pn$-$sd$IBM with
$F$-spin quantum number  has been studied by Yoshida et al \cite{Yo-09}.
Here, proton and neutron bosons are treated as the two components of a spin
$\spin$ boson and this spin is called $F$-spin. Yoshida et al found that
random interactions conserving $F$-spin generate  predominance of maximum
$F$-spin ($F_{max}$) ground states. It should be noted that the low-lying
states generated by $pn$-$sd$IBM correspond to those of $sd$IBM and all
$sd$IBM states will have $F=F_{max}$. Thus random interactions preserve the
property that the low-lying states generated by $pn$-$sd$IBM are those of
$sd$IBM. Similarly, using shell model with isospin conserving interactions
(here protons and neutrons correspond to the two projections of isospin
$\ct=\spin$), Kirson and Mizrahi \cite{Ki-07} showed that random
interactions generate natural isospin ordering. Denoting the lowest energy
state (les) for a given many nucleon isospin $T$ by $E_{les}(T)$, the
natural isospin ordering corresponds to $E_{les}(T_{min}) \leq
E_{les}(T_{min}+1) \leq \ldots$;  for even-even N=Z nuclei, $T_{min}=0$. 
Therefore, one can ask if  BEGOE(1+2)-$\cs$ generates a spin ordering.

As an application of BEGOE(1+2)-$\cs$, we present here results for the
probability of gs spin to be $S=S_{max}$ and also for natural spin ordering
(NSO). Here NSO corresponds to $E_{les}(S_{max}) \leq E_{les}(S_{max}-1)
\ldots$. In this analysis, we add the Majorana force or the space  exchange
operator to the Hamiltonian in Eq. (\ref{eq.begoe-s}). Note that $S$ in
BEGOE(1+2)-$\cs$ is similar to $F$-spin in the $pn$-$sd$IBM. First we will
derive the exchange interaction and then present some numerical results.

\subsection{$U(\Omega)$ algebra and space exchange operator}

In terms of boson creation ($b^\dagger$) and annihilation ($b$) operators,
the  sp states for $(\Omega)^m$ systems are $\l.\l| i,m_\cs \pm \spin\r.\ran =
b^\dagger_{i,\spin,m_\cs}\l|0\ran$ with
$i=1,2,\ldots,\Omega$. It can be easily identified that  the 4$\Omega^2$
number of one-body operators $A_{ij;\mu}^{r}$,
\be
A_{ij;\mu}^{r} =  \l(b_i^{\dagger}\tilde{b_j}\r)_{\mu}^{r}\;;\;\;\;\;
r=0,\;1\,,
\label{eq.maj1}
\ee
generate $U(2\Omega)$ algebra. In Eq. (\ref{eq.maj1}), $\tilde{b}_{i,\spin,
m_\cs}=(-1)^{\spin+m_\cs}  b_{i,\spin,-m_\cs}$.  The $U(2\Omega)$
irreducible representations  (irreps) are denoted trivially by the particle
number $m$ as they must be symmetric irreps $\{m\}$. The $\Omega^2$  number
of  operators $A_{ij}^{0}$ generate $U(\Omega)$ algebra and similarly there
is a $U(2)$ algebra generated by the number operator $\hat{n}$ and the spin
generators $S^1_\mu$,
\be
\barr{l}
\hat{n}=\dis\sqrt{2} \dis\sum_i A_{ii}^0\;;\;\;\;\;
S^1_\mu= \dis\frac{1}{\sqrt{2}} \dis\sum_i A_{ii;\mu}^{1}\;.
\earr \label{eq.ope}
\ee
Then we have the group-subgroup algebra $U(2\Omega) \supset U(\Omega)
\otimes SU(2)$ with $SU(2)$ generated by $S^1_\mu$. As the $U(2)$ irreps are
two-rowed, the $U(\Omega)$ irreps have to be two-rowed and they are labeled
by $\{m_1,m_2\}$ with $m=m_1+m_2$ and $S=(m_1-m_2)/2$; $m_1 \geq m_2 \geq
0$. Thus with respect to $U(\Omega) \otimes SU(2)$ algebra, many boson 
states are labeled by $\l| \{m_1,m_2\}, \xi \ran$ or equivalently by 
$\l| (m,S), \xi \ran$, where $\xi$ are extra labels required for a complete 
specification of the states. The quadratic Casimir  operator of the
$U(\Omega)$ algebra is,
\be
C_2[U(\Omega)] = 2\dis\sum_{i,j} A_{ij}^{0} \cdot A_{ji}^{0}
\label{eq.maj2}
\ee
and its eigenvalues are $\lan C_2[U(\Omega)] \ran^{\{m_1,m_2\}} = 
m_1(m_1+\Omega-1)+m_2(m_2+\Omega-3)$ or equivalently,
\be
\lan C_2[U(\Omega)] \ran^{(m,S)} = 
\dis\frac{m}{2}(2\Omega+m-4)+2S(S+1)\;.
\label{eq.maj4}
\ee
Note that the Casimir invariant of $SU(2)$ is $\hat{S}^2$ with eigenvalues
$S(S+1)$. Now we will show that the space exchange or the Majorana operator 
$\wm$ is simply related to $C_2\l[ U(\Omega)\r]$.

Majorana operator $\wm$ acting on a two-particle state exchanges the spatial
coordinates of the particles (index $i$) and leaves the  spin  quantum
numbers ($m_\cs$) unchanged. The operator form of $\wm$ is
\be
\wm = \dis\frac{\kappa}{2} \dis\sum_{i,j,m_\cs,m_\cs^\pr} \l(
b^\dagger_{j,m_\cs} b^\dagger_{i,m_\cs^\pr}\r) \l(
b^\dagger_{i,m_\cs} b^\dagger_{j,m_\cs^\pr}\r)^\dagger \;.
\label{eq.maj5}
\ee
Equation (\ref{eq.maj5}) gives, with $\kappa$ a constant,
\be
\wm = \dis\frac{\kappa}{2} \l\{ C_2\l[U(\Omega)\r] - \Omega \hat{n}\r\}\;.
\label{eq.maj6}
\ee
Then, combining Eqs. (\ref{eq.maj4}) and (\ref{eq.maj6}), we have
\be
\wm = \kappa \l\{\hat{n}\l( \dis\frac{\hat{n}}{4}-1\r)+\hat{S}^2\r\}\;.
\label{eq.maj7}
\ee
As seen from Eq. (\ref{eq.maj7}), exchange interaction with  $\kappa > 0$
generates gs with $S=S_{min}=0$($\spin$)  for even(odd) $m$ (this is
opposite to the result for fermion systems where the exchange interaction
generates gs with $S=S_{max}=m/2$ \cite{Ma-10c,JS-01}).  Now we will study
the interplay between random interactions and the Majorana force in
generating gs spin structure in boson systems. Note that for states with
boson number fixed, $\wm \propto \hat{S}^2$ as seen from Eq.
(\ref{eq.maj7}) and therefore, from now on, we refer to $\hat{S}^2$ as the
exchange interaction.

\subsection{Numerical results for $S_{max}=m/2$ ground states and natural
spin order}
\label{smax}

In order to understand the gs structure in BEGOE(1+2)-$\cs$, we have
studied $P(S=S_{max})$, the probability for the gs to be with spin
$S_{max}=m/2$, by adding the exchange term $\lambda_S\,S^2$ with $\lambda_S
> 0$ to the Hamiltonian in Eq. (\ref{eq.begoe-s}), i.e. using
\be 
\{H\}_{\mbox{BEGOE(1+2)-\cs}:\mbox{Exch}} = h(1) + \lambda\,\l[ \,
\{V^{s=0}(2)\} + \{V^{s=1}(2)\}\, \r]\, +\lambda_S\,S^2\;. 
\label{H-exch}
\ee 
Note that the operator $S^2$ is simple in the $(m,S)$ basis. Fig.
\ref{pofgs}a gives probability $P(S=S_{max})$ for the ground states to  have
spin $S=S_{max}$ as a function of exchange interaction strength $\lambda_S$
for $\lambda_0=\lambda_1=\lambda=0$, $0.1$, $0.2$, $0.3$ and $0.5$ and also
for $h(1)=0$ with $\lambda=1$. Similarly, Fig.
\ref{pofgs}b shows the results for NSO. Calculations are  carried out for 
($\Omega=4$, $m=10$) system using a 500 member ensemble and the  mean-field
Hamiltonian $h(1)$ is as defined in Section II.  

\subsubsection{Preponderance of $S_{max}=m/2$ ground states}

\begin{figure}
\includegraphics[width=4.5in,height=7in]{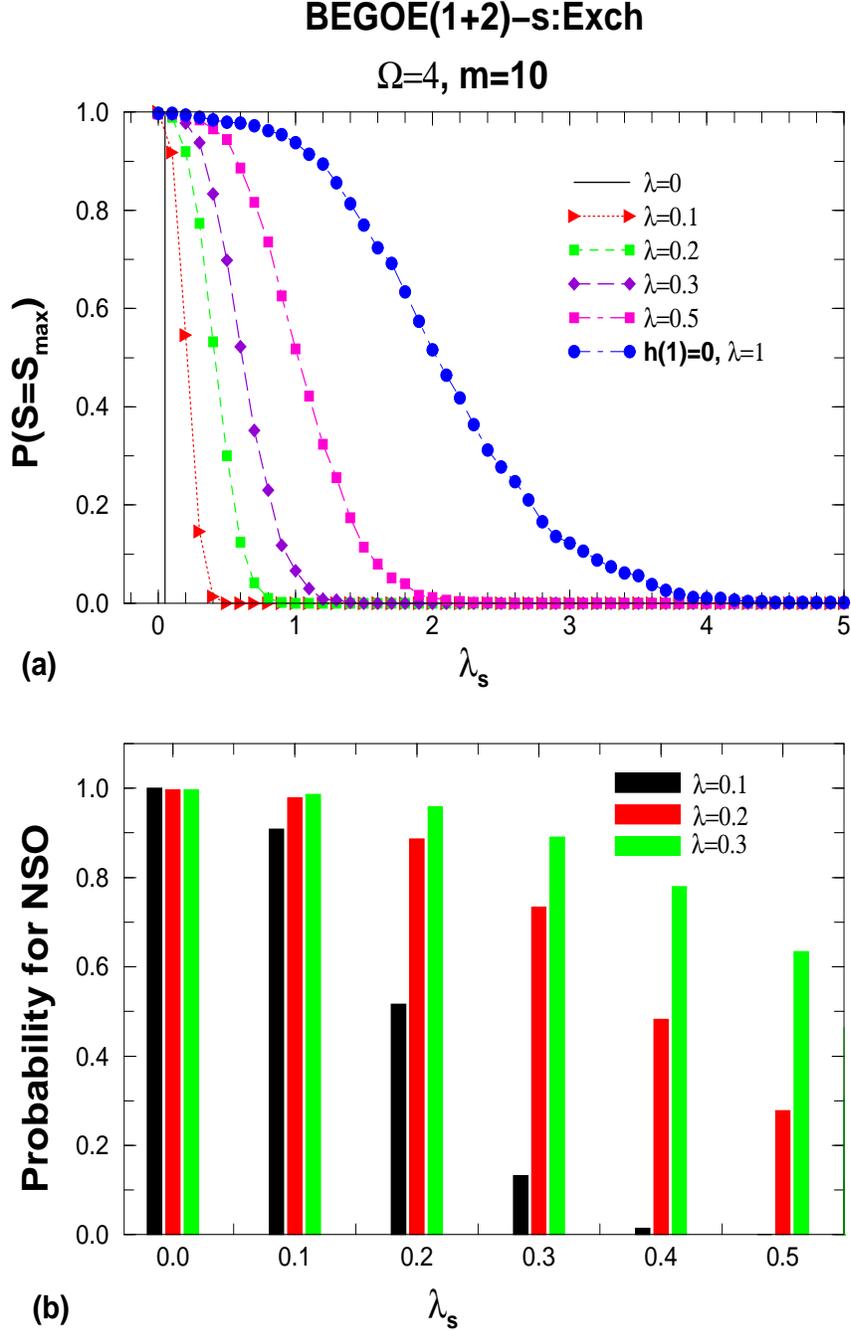}
\caption{(Color online) (a) Probability  for ground  states to have spin
$S=S_{max}$ as a function of the exchange interaction strength $\lambda_S
\geq 0$. (b) Probability for natural spin order (NSO)  as a function of
$\lambda_S$.  Results are shown for a 500 member 
BEGOE(1+2)-$\cs:\mbox{Exch}$ ensemble generated by Eq. (\ref{H-exch})  for a
system with $\Omega=4$ and $m=10$. Values of the interaction strength
$\lambda$ are shown in the figure.}
\label{pofgs}
\end{figure}

Let us begin with pure random two-body interactions. Then $h(1)=0$ in Eq.
(\ref{H-exch}). Now in the absence of the exchange interaction
($\lambda_S=0$), as seen from Fig. \ref{pofgs}a, ground states will have
$S=S_{max}$, i.e. the probability $P(S=S_{max}) = 1$. The variance
propagator (see Fig. \ref{prop}) derived earlier gives a  simple explanation
for this by applying the Jacquod and Stone prescription as discussed in Sec.
IV B. Thus pure random interactions generate preponderance of $S=S_{max}$
ground states. On the other hand, as discussed in Section V B,  the exchange
interaction acts in opposite direction by generating $S=S_{min}$ ground
states. Therefore, by adding the exchange interaction to the  $\{V(2)\}$
ensemble, $P(S=S_{max})$ starts decreasing as the strength $\lambda_S$
($\lambda_S > 0$) starts increasing. For the example considered in Fig.
\ref{pofgs}a, for $\lambda_S > 4$, we have $P(S=S_{max}) \sim 0$. The
complete variation with $\lambda_S$ is shown in Fig. \ref{pofgs}a marked
$h(1) = 0$ and $\lambda = 1$.

Similarly, on the other end, for $\lambda=0$ in Eq. (\ref{H-exch}), we have
$H=h(1)$ in the absence of the exchange interaction. In this situation,  as
all the bosons can occupy the lowest sp state, gs spin $S=S_{max}$.
Therefore, $P(S=S_{max})=1$. When the exchange interaction is turned on,
$P(S=S_{max})$ remains unity until  $\lambda_S$ equals the spacing between
the lowest two sp states divided by $m$. As in our example, the sp energies
are $\epsilon_i=i+1/i$, we have $P(S=S_{max}) = 1$ for $\lambda_S < 0.05$. 
Then  $P(S=S_{max})$ drops to zero for $\lambda_S \geq 0.05$. This variation
with $\lambda_S$ is shown in Fig. \ref{pofgs}a marked  $\lambda = 0$. Figure
\ref{pofgs}a also shows the variation of $P(S=S_{max})$ with $\lambda_S$ for
several values of $\lambda$ between $0.1$ and $0.5$. It is seen that there
is a critical value ($\lambda_S^c$) of $\lambda_S$ after which $P(S=S_{max})
= 0$ and its value increases with $\lambda$. Also, the variation of
$P(S=S_{max})$ with $\lambda_S$ becomes slower as $\lambda$ increases.

In summary, results in Fig. \ref{pofgs}a  clearly show that with random
interactions there is preponderance of $S=S_{max}=m/2$ ground states. This
is unlike for fermions where there is preponderance of $S=S_{min}=0(\spin)$
ground states for $m$ even(odd). With the addition of the exchange 
interaction,  $P(S=S_{max})$ decreases and finally goes to zero for 
$\lambda_S \geq  \lambda_S^c$ and the value of $\lambda_S^c$ increases with
$\lambda$. We have also carried out calculations for ($\Omega=4$, $m=11$)
system using a 100 member ensemble  and the results are close to those given
in Fig. \ref{pofgs}a.  All these explain the results given in \cite{Yo-09}
where random interactions are employed within $pn$-$sd$IBM.

\subsubsection{Natural spin ordering}

For the system considered in  Fig. \ref{pofgs}a, for each member of the
ensemble, eigenvalue of the  lowest state for each spin $S$ is calculated
and using these, we have obtained total number of members $N_\lambda$ having
NSO as a function of  $\lambda_S$ for $\lambda=0.1, 0.2$ and $0.3$ using the
Hamiltonian given in Eq. (\ref{H-exch}).  As stated in Section V A, the NSO
here corresponds to (as $S=S_{max}$ is the spin of the gs of the system)
$E_{les}(S_{max}) < E_{les}(S_{max}-1) < E_{les}(S_{max}-2) < \ldots$.  The
probability for NSO is $N_\lambda/500$ and the results are shown in Fig.
\ref{pofgs}b.  In the absence of the exchange interaction, as seen from Fig.
\ref{pofgs}b, NSO is found in all the members  independent of $\lambda$.
Thus random interactions strongly favor NSO. The presence of exchange
interaction reduces the probability for NSO. Comparing Figs. \ref{pofgs}a
and \ref{pofgs}b,  it is clearly seen that with increasing exchange
interaction strength, probability for gs state spin  to be  $S=S_{max}$ is
preserved for much larger values of $\lambda_S$ (with a fixed $\lambda$)
compared to the NSO. Therefore for preserving both $S=S_{max}$ gs and the
NSO with high probability, the $\lambda_S$ value has to be small. We have
also verified this for the ($\Omega=4$, $m=11$) system. Finally, it is
plausible to argue  that the  results in Fig. \ref{pofgs}  obtained using
BEGOE(1+2)-$\cs$ are generic for boson systems with spin.  Now we will turn
to pairing in BEGOE(2)-$\cs$.

\section{Pairing in BEGOE(2)-$\cs$}

Pairing correlations are known to be important not only for fermion systems 
but also for boson systems \cite{Pe-10}. An important issue that is raised
in the recent years is: to what extent random interactions carry features of
pairing. See \cite{Ma-10,Zh-04,Zel-04,Ho-07} for some results for fermion
systems. In order to address this question for boson systems, first we will
identify the pairing algebra in $(\Omega,m,S)$ spaces of BEGOE(2)-$\cs$.
Then we will consider expectation values of the pairing Hamiltonian in the
eigenstates  generated by BEGOE(2)-$\cs$ as they carry signatures of
pairing.

\subsection{$U(2\Omega)\supset [U(\Omega) \supset SO(\Omega)] \otimes 
SU_S(2)$ Pairing symmetry}

In constructing BEGOE(2)-$\cs$, it is assumed that spin is a good symmetry
and thus the $m$-particle states carry spin ($S$) quantum number.  Now,
following the $SO(5)$ pairing algebra for fermions \cite{Fl-64}, it is
possible to consider pairs that are vectors in spin space. The pair creation
operators $P_{i:\mu}$ for the level $i$ and the generalized  pair creation
operators (over the $\Omega$ levels) $P_\mu$, with $\mu=-1,0,1$, in spin
coupled representation, are
\be
P_\mu =  \dis\frac{1}{\sqrt{2}} \dis\sum_i \l(b^\dagger_i 
b^\dagger_i\r)^1_\mu = \dis\sum_i P_{i:\mu}\;,\;\;\;
\l( P_\mu \r)^{\dagger}=\dis\frac{1}{\sqrt{2}}
\dis\sum_i(-1)^{1-\mu}\l(\tilde{b_i}\tilde{b_i}\r)^1_{-\mu}\;.
\label{eq.npa1}
\ee
Therefore in the space defining BEGOE(2)-$s$, the  pairing Hamiltonian $H_p$
and its two-particle matrix elements are,
\be
H_p = \dis\sum_\mu P_\mu \l( P_\mu \r)^\dagger\;,\;\;\;\lan (k\ell) s 
\mid H_p \mid (ij) s \ran = \delta_{s,1}\,\delta_{i,j} \, \delta_{k,\ell}\;.
\label{eq.npa2}
\ee
With this, we will proceed to identify and analyze the pairing algebra. It
is easy to verify that the  $\Omega(\Omega-1)/2$ number of operators
$C_{ij}=A_{ij}^{0}-A_{ji}^{0}$, $i>j$ generate a $SO(\Omega)$ subalgebra of
the $U(\Omega)$ algebra. Therefore we have $U(2\Omega) \supset [U(\Omega)
\supset SO(\Omega)] \otimes SU(2)$. We will show that the irreps of
$SO(\Omega)$ algebra are uniquely labeled by the seniority quantum number
$v$ and a reduced spin $\tilde{s}$  similar to the reduced isospin
introduced in the context of nuclear
shell model \cite{Fl-52} and they in turn define the eigenvalues of $H_p$.
The quadratic Casimir operator of the $SO(\Omega)$ algebra is,
\be
C_2[SO(\Omega)] = 2\dis\sum_{i>j} C_{ij} \cdot C_{ji} \;.
\label{eq.npa5}
\ee
Carrying out angular momentum algebra \cite{Ed-77} it can be shown that,
\be
C_2[SO(\Omega)] = C_2[U(\Omega)] - 2\,H_p - \hat{n}\;.
\label{eq.npa6}
\ee
The quadratic Casimir operator of the $U(\Omega)$ algebra is given in 
Eq. (\ref{eq.maj2}). Before discussing the eigenvalues of the pairing
Hamiltonian $H_p$, let us first consider the irreps of $SO(\Omega)$.

Given the two-rowed $U(\Omega)$ irreps $\{m_1,m_2\}$; $m_1+m_2=m$, 
$m_1-m_2=2S$, it should be clear that the $SO(\Omega)$ irreps should be of 
$[v_1,v_2]$ type and for later simplicity we use $v_1+v_2=v$ and 
$v_1-v_2=2\tilde{s}$. The quantum number $v$ is called seniority and
$\tilde{s}$ is called reduced spin. The $SO(\Omega)$ irreps for a given
$\{m_1,m_2\}$ can be  obtained as follows. First expand the $U(\Omega)$
irrep $\{m_1,m_2\}$ in  terms of totally symmetric irreps, 
\be
\{m_1,m_2\}=\{m_1\} \times \{m_2\}-\{m_1+1\} \times \{m_2-1\}\;. 
\label{eq.npa22}
\ee
Note that the irrep multiplication in Eq. (\ref{eq.npa22}) is a Kronecker
multiplication \cite{Ko-06b,Wy-70}.
For a totally symmetric $U(\Omega)$ irrep $\{m^\pr\}$, the $SO(\Omega)$ 
irreps are given by the well-known result
\be
\{m^\pr\} \to [v] = [m^\pr] \oplus [m^\pr-2] \oplus \ldots \oplus [0] 
\mbox{ or } [1]\;. 
\label{eq.npa23}
\ee
Finally, reduction of the Kronecker product of two symmetric $SO(\Omega)$
irreps $[v_1]$ and $[v_2]$, $\Omega>3$ into $SO(\Omega)$ irreps $[v_1,v_2]$
is given by (for $v_1 \geq v_2$) \cite{Ko-06b,Wy-70},
\be
[v_1] \times [v_2] = \dis\sum_{k=0}^{v_2}\dis\sum_{r=0}^{v_2-k}
[v_1-v_2+k+2r,k] \oplus \;.
\label{eq.npa10}
\ee
Combining Eqs. (\ref{eq.npa22}), (\ref{eq.npa23}) and (\ref{eq.npa10}) gives
the $\{m_1,m_2\} \to [v_1,v_2]$ reductions. It is easy to implement this
procedure on a computer. 

Given the space defined by $\l| \{m_1,m_2\}, [v_1,v_2], \alpha \ran$, with
$\alpha$ denoting extra labels needed for a complete specification of the
state, the eigenvalues of $C_2[SO(\Omega)]$ are \cite{Ko-06b}
\be
\lan C_2[SO(\Omega)] \ran^{\{m_1,m_2\}, [v_1,v_2]} = 
v_1(v_1+\Omega-2)+v_2(v_2+\Omega-4)\;. 
\label{eq.npa31}
\ee
Now changing $\{m_1,m_2\}$ to $(m,S)$ and $[v_1,v_2]$ to $(v,\tilde{s})$ and
using Eqs. (\ref{eq.npa6}) and (\ref{eq.maj4}) will give the formula for the
eigenvalues of the  pairing Hamiltonian $H_p$. The final result is,
\be
E_p(m,S,v,\tilde{s}) = \lan H_p \ran^{m,S,v,\tilde{s}} = 
\dis\frac{1}{4}(m-v)(2\Omega-6+m+v) + [S(S+1)-\tilde{s}(\tilde{s}+1)]\;.
\label{eq.npa8}
\ee
This is same as the result that follows from Eq. (18) of \cite{Fl-64} for
fermions by using $\Omega \to -\Omega$ symmetry. From now on, we denote the
$U(\Omega)$ irreps by $(m,S)$ and $SO(\Omega)$ irreps by $(v,\tilde{s})$. In
Table \ref{red}, for $(\Omega,m) = (4,10),(5,8)$ and $(6,6)$ systems, given
are the $(m,S) \to (v,\tilde{s})$ reductions, the pairing eigenvalues given
by Eq. (\ref{eq.npa8}) in the spaces defined by these irreps and also the
dimensions of the  $U(\Omega)$ and $SO(\Omega)$ irreps. The dimensions 
$d(\Omega,m,S)$ of the $U(\Omega)$ irreps $(m,S)$ are given by Eq.
(\ref{eq.msdim}). Similarly, the dimension $\cads(v_1,v_2) \Leftrightarrow 
\cads(v,\tilde{s})$ of the $SO(\Omega)$ irreps  $[v_1,v_2]$ follow from Eqs.
(\ref{eq.npa23}) and (\ref{eq.npa10}) and they will give
\be
\barr{l}
\cads(v_1,v_2) =  \cads(v_1) \cads(v_2) - \dis\sum_{k=0}^{v_2-1} 
\dis\sum_{r=0}^{v_2-k} \cads(v_1-v_2+k+2r,k)\;;\\ \\
\cads(v) = \dis\binom{\Omega+v-1}{v} - \dis\binom{\Omega+v-3}{v-2}\;.
\earr \label{eq.dim}
\ee
Note that in general the $SO(\Omega)$ irreps $(v,\tilde{s})$ can appear more
than once in the reduction of $U(\Omega)$ irreps $(m,S)$. For example,
$(2,1)$ irrep of  $SO(\Omega)$ appears twice in the reduction of the
$U(\Omega)$ irrep $(10,1)$.   

It is useful to remark that just as the fermionic $SO(5)$ pairing algebra
for nucleons in $j$ orbits \cite{Pa-65,He-65,Fl-64}, there will be a
$SO(4,1)$ complementary pairing algebra corresponding to the  $SO(\Omega)$
subalgebra. The ten operators $P^1_\mu$,  $(P^1_\mu)^\dagger$, $S^1_\mu$ and
$\hat{n}$ form the $SO(4,1)$ algebra. It is possible to exploit this algebra
to derive properties of the eigenstates defined by the pairing Hamiltonian
but this will be discussed elsewhere. 

\subsection{Pairing expectation values}

\begin{figure}[htp]
\includegraphics[width=6in]{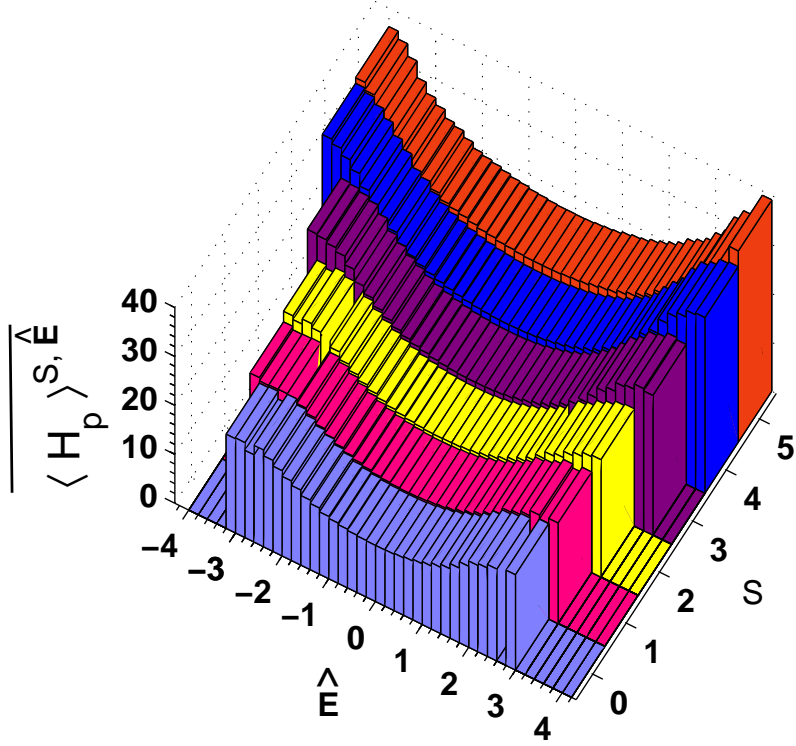}
\caption{(Color online) Ensemble averaged pairing expectation values
$\overline{\lan H_p \ran^{S,\widehat{\bee}}}$ vs $\widehat{\bee}$ and $S$,
shown as a 3D  histogram, for a 500 member BEGOE(2)-$\cs$ ensemble with
$\Omega=4$ and  $m=10$. The bin-size is 0.2 for $\widehat{\bee}$. Note that
the $\widehat{\bee}$ label in this figure is different from the
$\widehat{E}$ used in Figs. \ref{den} and \ref{n4m11}a.}
\label{pair}
\end{figure}

Pairing expectation values are defined by $\lan H_p \ran^{S,E} =  \lan m,S,E
\mid H_p \mid m,S,E \ran$ for eigenstates with energy $E$ and spin $S$
generated by a Hamiltonian $H$ for a system of $m$ bosons in $\Omega$ number
of sp orbitals (for simplicity, we have dropped $\Omega$ and $m$ labels in
$\lan H_p \ran^{S,E}$). In our analysis, $H$ is a member of BEGOE(2)-$\cs$.
As we will be comparing the results for all spins at a given energy $E$, for
each member of the ensemble the eigenvalues for all spins are zero centered
and normalized using the $m$-particle energy centroid $E_c(m)=\lan H \ran^m$
and spectrum width $\sigma(m)=[\lan H^2 \ran^m -\{E_c(m)\}^2]^{1/2}$.
Then the eigenvalues $E$ for all $S$ are changed to 
$\widehat{\bee}=[E-E_c(m)]/\sigma(m)$. Using the method described in Section
II, the $H_p$ matrix is constructed in good $M_S$ basis and transformed into
the  eigenbasis of a given $S$  for each member of the  BEGOE(2)-$\cs$
ensemble. Then the ensemble average of the diagonal elements of the $H_p$
matrix will give the ensemble averaged pairing expectation values
$\overline{\lan H_p \ran^{S,E}} \Leftrightarrow  \overline{\lan H_p
\ran^{S,\widehat{\bee}}}$.  Using this procedure for a  500 member
BEGOE(2)-$\cs$ ensemble with $\Omega=4$, $m=10$ and $S=0-5$, results for 
$\overline{\lan H_p \ran^{S,\widehat{\bee}}}$ as a  function of energy
$\widehat{\bee}$ (with $\widehat{\bee}$ as described above)  and spin $S$
are obtained and they are shown  as a 3D histogram in Fig. \ref{pair}. From
Table \ref{red}, it is seen that the maximum value of the eigenvalues
$E_p(m,S,v,\tilde{s})$ increases with spin $S$ for a fixed-$(\Omega,m)$. 
The values are $28$, $32$, $34$, $42$, $48$ and $60$ for $S=0-5$
respectively for $\Omega=4$ and $m=10$.  Numerical results in Fig.
\ref{pair} also  show that for states near the lowest $\widehat{\bee}$
value,  $\overline{\lan H_p \ran^{S,\widehat{\bee}}}$ increases  with spin 
$S$. Thus random interactions preserve this property of the pairing
Hamiltonian in addition to generating $S=S_{max}$ ground states as discussed
in  Section \ref{smax}. It is useful to remark that random interactions will
not generate $S=S_{max}$ ground states with $(v,\tilde{s})=(m,m/2)$ as
required for example in the $pn$-$sd$IBM. This needs explicit inclusion of
pairing and exchange terms in the Hamiltonians defined by Eqs.
(\ref{eq.begoe-s})  and (\ref{eq.v2}).

For a given spin $S$, the pairing expectation values as a function of $E$
are expected, for two-body ensembles,  to be given by a ratio of expectation
value density (EVD) Gaussian (the first two moments given by $\lan H_p H
\ran^{m,S}$ and $\lan H_p H^2 \ran^{m,S}$) and the eigenvalue density 
Gaussian with normalization given by $\lan H_p \ran^{m,S}$ and this itself 
will be a Gaussian \cite{Ma-09}.  Let us denote the EVD centroid by
$E_c(m,S:H_p)$ and width by $\sigma(m,S:H_p)$. Then  the ratio of Gaussians
will give 
\be 
\barr{l} 
\overline{\lan H_p
\ran^{S,\widehat{E}}} = \dis\frac{\lan H_p \ran^{m,S}}
{\widehat{\sigma}(m,S)} \exp{\dis\frac{\widehat{\epsilon}^2(m,S)}
{2\l[1-\widehat{\sigma}^2(m,S)\r]}} \;
\exp\l\{\dis\frac{(\widehat{\sigma}^2(m,S)-1)}{2\widehat{\sigma}^2(m,S)} \l[
\widehat{E} - \dis\frac{\widehat{\epsilon}(m,S)}
{1-\widehat{\sigma}^2(m,S)}\r]^2\r\}\;. 
\earr \label{eq.ratio} 
\ee 
Here, $\widehat{\epsilon}(m,S)=\{ E_c(m,S:H_p)-E_c(m,S)\}/
\sigma(m,S)$, $\widehat{\sigma}(m,S) = \sigma(m,S:H_p)/\sigma(m,S)$ and 
$\widehat{E}=[\sigma(m)/\sigma(m,S)]\{\widehat{\bee}-\ce\}$;
$\ce=[E_c(m,S)-E_c(m)]/\sigma(m)$. The
Gaussian form given by Eq. (\ref{eq.ratio}) is clearly  seen in Fig.
\ref{pair} and this also gives a quantitative description of the results.
Note that in our example, $\widehat{\epsilon}(10,S) = 0.001$, $0.001$,
$0.001$, $0.002$, $0.002$, $0.003$ and $\widehat{\sigma}(10,S) = 1.045$,
$1.047$, $1.053$, $1.062$, $1.073$, $1.082$ respectively for $S=0-5$.

\section{Conclusions}

In the present work, we have introduced the BEGOE(1+2)-$\cs$ ensemble and a
method for constructing BEGOE(1+2)-$\cs$ for numerical calculations has been
described. Numerical examples are used to show that, like the spinless
BEGOE(1+2), the spin BEGOE(1+2)-$\cs$ ensemble also generates Gaussian
density of  states in the dense limit. Similarly, BEGOE(2)-$\cs$ exhibits
GOE level  fluctuations. On the other hand, BEGOE(1+2)-$\cs$ exhibits
Poisson to GOE transition as the interaction strength $\lambda$ is increased
and the transition marker $\lambda_c$ is found to decrease with increasing
spin. Moreover, ensemble averaged covariances in energy centroids and
spectral variances for BEGOE(2)-$\cs$ between spectra with different
particle numbers and spins are studied using the propagation formulas
derived for the  energy centroids and spectral variances. For $\Omega=12$
systems, the cross correlations in energy centroids are  $\sim 15$\% and
they reduce to $\sim 4$\% for spectral variances. We have also derived the
exact formula for the ensemble averaged fixed-$(m,S)$ spectral variances and
demonstrated that the variance propagator gives a simple explanation for the
preponderance of spin $S=S_{max}$ ground states generated by random
interactions as in $pn$-$sd$IBM. It is also shown, by including exchange
interaction $\hat{S}^2$ in BEGOE(1+2)-$\cs$, that random interactions
preserving spin symmetry strongly favor NSO (just  as with isospin in 
nuclear shell model). In addition, we have identified the pairing
$SO(\Omega)$ symmetry and showed using numerical examples that random
interactions exhibit pairing correlations in the gs region and also they
generate a Gaussian form for the variation of the pairing expectation values
with respect to energy. 

Results in this paper represent a beginning in systematic studies of
BEGOE(1+2)-$\cs$ ensemble. To go beyond the present work requires: (i)
dealing with systems with much larger $\Omega$ and $m$ values and then the
matrix dimensions will be $10^5-10^6$ and higher;  (ii) further group
theoretical analysis for deriving higher moments of the two-point function
and this requires new results for Wigner and Racah coefficients for general
$U(\Omega)$ algebras (see for example \cite{Ma-10b}); and (iii) identifying
and analyzing measures that can be used in experimental data analysis, say
for example data from ultracold gases. With progress in (i) and (ii), it is
possible to investigate more systematically BEGOE(1+2)-$\cs$ as a function
of the strength parameter $\lambda$ by analyzing the delocalization and
other related measures (see for example \cite{Ma-10,Sa-10,Sa-10a}). With
these it is possible to address (iii) but this is for future.

Further extensions of BEGOE(1+2)-$\cs$ including $\cs=1$, $2$, $\ldots$
degrees of freedom for bosons, as emphasized in the Introduction, are
relevant for spinor BEC studies \cite{Pe-10,Yi-07}. These extended BEGOE's
will be explored in future. Finally, it is useful to mention that the
numerical code developed  for constructing BEGOE(1+2)-$\cs$ ensemble in
fixed-$(m,S)$ spaces can be used in analyzing ensembles generated by
Hamiltonians preserving only $M_S$. Here it is important to point out that,
besides the $SO(\Omega)$ pairing discussed in Section VI, there is another
pairing algebra that corresponds to $U(2\Omega) \supset  SO(2\Omega)
\supset  [SO(\Omega)\otimes SO(2)]$ as pointed out in \cite{Ko-06b}. The 
pairing Hamiltonian corresponding to this group-subgroup chain preserves
only $M_S$  but not $S$ and hence, relevant for BEGOE(1+2) with fixed $M_S$.

\acknowledgments

All calculations in this paper have been carried out on the HPC cluster
facility at Physical Research Laboratory and the DELL workstation at MSU,
Baroda.

\renewcommand{\theequation}{A\arabic{equation}}
\setcounter{equation}{0}   
\section*{APPENDIX A}  

Let us consider a system of $m$ spinless bosons occupying $N$ sp states $\l|
\l. \nu_i \ran \r.$, $i=1,2,\ldots,N$ and the Hamiltonian is say two-body.
Then the Hamiltonian operator is 
\be
\widehat{H} = \dis\sum_{\nu_i \leq \nu_j,\;\nu_k \leq \nu_l}
\dis\frac{ \lan \nu_k\;\nu_l
\mid H \mid \nu_i\;\nu_j\ran}{\dis\sqrt{(1+\delta_{ij})
(1+\delta_{kl})}}\;b^\dg_{\nu_k}\,b^\dg_{\nu_l}\,
b_{\nu_i}\,b_{\nu_j}\;.
\ee
with the symmetries for the symmetrized two-body matrix elements 
$\lan \nu_k\;\nu_l \mid H \mid \nu_i\;\nu_j\ran$ being,
\be
\barr{c}
\lan \nu_k\;\nu_l \mid H \mid \nu_j\;\nu_i\ran = \lan \nu_k\;\nu_l
\mid H \mid \nu_i\;\nu_j\ran \;,\\
\lan \nu_k\;\nu_l \mid H \mid \nu_i\;\nu_j\ran = \lan \nu_i\;\nu_j
\mid H \mid \nu_k\;\nu_l\ran \;.\\
\earr \label{eq.app1}
\ee
The Hamiltonian matrix $H(m)$ in $m$-particle spaces contains three
different types of non-zero matrix elements and explicit formulas for these
are \cite{PDPK},
\be
\barr{l}
\lan \dis\prod _{r=i,j,\ldots} \l(\nu_r\r)^{n_r} \mid H \mid 
\dis\prod _{r=i,j,\ldots} \l(\nu_r\r)^{n_r} \ran = 
\dis\sum_{i \geq j }\; \dis\frac{n_i\l(n_j-\delta_{ij}\r)}{\l(1+
\delta_{ij}\r)} \; \lan \nu_i
\nu_j \mid H \mid \nu_i \nu_j \ran \;,\\
\\
\lan \l(\nu_i\r)^{n_i-1} \l(\nu_j\r)^{n_j+1} 
\dis\prod_{r^\prime=k,l,\ldots} 
\l(\nu_{r^\prime}\r)^{n_{r^\prime}} \mid H \mid 
\dis\prod _{r=i,j,\ldots} \l(\nu_r\r)^{n_r} \ran = \\ \\
\dis\sum_{k^\prime}\; \l[ \dis\frac{n_i\l(n_j+1\r)\l(n_{k^\prime}-
\delta_{k^\prime i}\r)^2}{\l(1+\delta_{k^\prime i}\r) \l(1+\delta_{
k^\prime j}\r)}\r]^{1/2} \; \lan \nu_{k^\prime} \nu_j \mid H \mid
\nu_{k^\prime} \nu_i \ran \;,\\
\\
\lan \l(\nu_i\r)^{n_i+1} \l(\nu_j\r)^{n_j+1} \l(\nu_k\r)^{n_k-1}
\l(\nu_l\r)^{n_l-1} \dis\prod _{r^\pr=m,n,\ldots} 
\l(\nu_{r^\pr}\r)^{n_{r^\pr}} \mid H \mid 
\dis\prod _{r=i,j,\ldots} \l(\nu_r\r)^{n_r} \ran = \\ \\
\l[ \dis\frac{n_k\l(n_l-\delta_{kl}\r)\l(n_i+1\r)\l(n_j+1+
\delta_{ij}\r)}{\l(1+\delta_{ij}\r) \l(1+\delta_{kl}\r)} \r]^{1/2}\; 
\lan \nu_i \nu_j \mid H \mid \nu_k \nu_l \ran \;.
\earr \label{eq.app2}
\ee
In the second equation in Eq. (\ref{eq.app2}), $i \neq j$ and in the third
equation, four combinations are possible: (i) $k=l$, $i=j$, $k \neq i$; (ii)
$k=l$, $i \neq j$, $k \neq i$, $k \neq j$; (iii) $k \neq l$, $i=j$, $i \neq
k$, $i \neq l$; and (iv) $i \neq j \neq k \neq l$.  BEGOE(2) for spinless
boson systems is defined by Eqs. (\ref{eq.app1}) and (\ref{eq.app2}) with
the $H$ matrix in two-particle spaces being GOE.  Note that the $H(m)$
matrix dimension is $\binom{N+m-1}{m}$ and the number of independent matrix
elements is $d_2(d_2+1)/2$ where $d_2 = N(N+1)/2$. It is useful to mention
that the formulas for the energy centroids and spectral variances in
$m$-boson spaces  follow from Eqs. (\ref{eq.bcent}), (\ref{eq.bvar}),
(\ref{eq.bvar1}) and (\ref{eq.bvar2}) with $S=m/2$, putting the $s=0$ matrix
elements to zero and replacing $\Omega$ by $N$.

\newpage

\setlength{\LTcapwidth}{7in}

\begin{longtable}{cccccccccc}
\caption{Classification of states in the $U(2\Omega)\supset$  $[U(\Omega)
\supset  SO(\Omega)]\otimes SU_S(2)$ limit for $(\Omega,m) = (4,10),(5,8)$
and $(6,6)$.  Given are $U(\Omega)$ labels $(m, S)$ and $SO(\Omega)$ labels 
$(v, \tilde{s})$ with the corresponding dimensions  $d(\Omega,m,S)$ and 
$\cads(v,\tilde{s})$ respectively and also the pairing eigenvalues  $E_p(m,
S,v,\tilde{s})$.  Note that $\sum_{v,\tilde{s}} r 
\cads(v,\tilde{s})=d(\Omega,m,S)$; here $r$ denotes multiplicity of the 
$SO(\Omega)$ irreps and in the table,  they are shown only for the cases
when $r>1$.} \\

\hline 
$\Omega$ & $m$ & $(m,S)_{d(\Omega,m,S)}$ &
$(v,\tilde{s})^r_{\cads(v,\tilde{s})}$ & $E_p(m, S,v,\tilde{s})$ & $\Omega$
& $m$ & $(m,S)_{d(\Omega,m,S)}$ &  $(v,\tilde{s})^r_{\cads(v,\tilde{s})}$ &
$E_p(m, S,v,\tilde{s})$ 
\\ 
\hline 
\endfirsthead

\multicolumn{10}{c}%
{{\bfseries \tablename\ \thetable{} -- continued}} \\
\hline
$\Omega$ & $m$ & $(m,S)_{d(\Omega,m,S)}$ &
$(v,\tilde{s})^r_{\cads(v,\tilde{s})}$ & $E_p(m, S,v,\tilde{s})$ & $\Omega$
& $m$ & $(m,S)_{d(\Omega,m,S)}$ &  $(v,\tilde{s})^r_{\cads(v,\tilde{s})}$ &
$E_p(m, S,v,\tilde{s})$  \\ \hline 
\endhead

\hline
\endfoot

\hline \hline
\endlastfoot
$4$ & $10$ & $( 10 ,  0 )_{196}$ & $( 2 ,  0 )_{6}$ & $28$ &
$5$ & $8$ & $( 8 ,  0 )_{490}$ & $( 0 ,  0 )_{1}$ & $24$ \\     	
 &  &  & $( 4 ,  1 )_{30}$ & $22$ &  
 &  &  & $( 2 ,  1 )_{14}$ & $19$ \\     	
 &  &  & $( 6 ,  2 )_{70}$ & $12$ &  
 &  &  & $( 4 ,  2 )_{55}$ & $10$ \\     	
 &  &  & $( 6 ,  0 )_{14}$ & $18$ &  
 &  &  & $( 4 ,  0 )_{35}$ & $16$ \\     	
 &  &  & $( 8 ,  1 )_{54}$ & $8$ &  
 &  &  & $( 6 ,  1 )_{220}$ & $7$ \\     	
 &  &  & $( 10 ,  0 )_{22}$ & $0$ &  
 &  &  & $( 8 ,  0 )_{165}$ & $0$ \\     	
 &  & $( 10 ,  1 )_{540}$ & $( 2 ,  1 )^ 2_{9}$ & $28$ &  
 &  & $( 8 ,  1 )_{1260}$ & $( 2 ,  1 )_{14}$ & $21$ \\     	
 &  &  & $( 4 ,  2 )^ 2_{25}$ & $20$ &  
 &  &  & $( 4 ,  2 )_{55}$ & $12$ \\     	
 &  &  & $( 6 ,  3 )_{49}$ & $8$ &  
 &  &  & $( 4 ,  1 )^ 2_{81}$ & $16$ \\     	
 &  &  & $( 4 ,  1 )_{30}$ & $24$ &  
 &  &  & $( 6 ,  2 )_{260}$ & $5$ \\     	
 &  &  & $( 6 ,  2 )_{70}$ & $14$ &  
 &  &  & $( 6 ,  1 )_{220}$ & $9$ \\     	
 &  &  & $( 6 ,  1 )^ 2_{42}$ & $18$ &  
 &  &  & $( 8 ,  1 )_{455}$ & $0$ \\     	
 &  &  & $( 8 ,  2 )_{90}$ & $6$ &  
 &  &  & $( 2 ,  0 )_{10}$ & $23$ \\     	
 &  &  & $( 8 ,  1 )_{54}$ & $10$ &  
 &  &  & $( 6 ,  0 )_{84}$ & $11$ \\     	
 &  &  & $( 10 ,  1 )_{66}$ & $0$ &  
 &  & $( 8 ,  2 )_{1500}$ & $( 4 ,  2 )^ 2_{55}$ & $16$ \\     	
 &  &  & $( 0 ,  0 )_{1}$ & $32$ &  
 &  &  & $( 6 ,  3 )_{140}$ & $3$ \\     	
 &  &  & $( 4 ,  0 )_{10}$ & $26$ &  
 &  &  & $( 6 ,  2 )_{260}$ & $9$ \\     	
 &  &  & $( 8 ,  0 )_{18}$ & $12$ &  
 &  &  & $( 8 ,  2 )_{625}$ & $0$ \\     	
 &  & $( 10 ,  2 )_{750}$ & $( 4 ,  2 )^ 2_{25}$ & $24$ &  
 &  &  & $( 2 ,  1 )^ 2_{14}$ & $25$ \\     	
 &  &  & $( 6 ,  3 )_{49}$ & $12$ &  
 &  &  & $( 4 ,  1 )_{81}$ & $20$ \\     	
 &  &  & $( 6 ,  2 )^ 2_{70}$ & $18$ &  
 &  &  & $( 6 ,  1 )_{220}$ & $13$ \\     	
 &  &  & $( 8 ,  3 )_{126}$ & $4$ &  
 &  &  & $( 0 ,  0 )_{1}$ & $30$ \\     	
 &  &  & $( 8 ,  2 )_{90}$ & $10$ &  
 &  &  & $( 4 ,  0 )_{35}$ & $22$ \\     	
 &  &  & $( 10 ,  2 )_{110}$ & $0$ &  
 &  & $( 8 ,  3 )_{1155}$ & $( 6 ,  3 )_{140}$ & $9$ \\     	
 &  &  & $( 2 ,  1 )_{9}$ & $32$ &  
 &  &  & $( 8 ,  3 )_{595}$ & $0$ \\     	
 &  &  & $( 4 ,  1 )^2_{30}$ & $28$ &  
 &  &  & $( 4 ,  2 )_{55}$ & $22$ \\     	
 &  &  & $( 6 ,  1 )_{42}$ & $22$ &  
 &  &  & $( 6 ,  2 )_{260}$ & $15$ \\     	
 &  &  & $( 8 ,  1 )_{54}$ & $14$ &  
 &  &  & $( 2 ,  1 )_{14}$ & $31$ \\     	
 &  &  & $( 2 ,  0 )_{6}$ & $34$ &  
 &  &  & $( 4 ,  1 )_{81}$ & $26$ \\     	
 &  &  & $( 6 ,  0 )_{14}$ & $24$ &  
 &  &  & $( 2 ,  0 )_{10}$ & $33$ \\     	
 &  & $( 10 ,  3 )_{770}$ & $( 6 ,  3 )^ 2_{49}$ & $18$ &  
 &  & $( 8 ,  4 )_{495}$ & $( 8 ,  4  )_{285}$ & $0$ \\     	
 &  &  & $( 8 ,  4 )_{81}$ & $2$ &  
 &  &  & $( 6 ,  3  )_{140}$ & $17$ \\     	
 &  &  & $( 8 ,  3 )_{126}$ & $10$ &  
 &  &  & $( 4 ,  2  )_{55}$ & $30$ \\     	
 &  &  & $( 10 ,  3 )_{154}$ & $0$ &  
 &  &  & $( 2 ,  1  )_{14}$ & $39$ \\     	
 &  &  & $( 4 ,  2 )^ 2_{25}$ & $30$ &  
 &  &  & $( 0 ,  0  )_{1}$ & $44$ \\     	
 &  &  & $( 6 ,  2 )_{70}$ & $24$ &  
$6$ & $6$ & $( 6 ,  0 )_{490}$ & $( 2 ,  0 )_{15}$ & $14$ \\     	
 &  &  & $( 8 ,  2 )_{90}$ & $16$ &  
 &  &  & $( 4 ,  1 )_{175}$ & $6$ \\     	
 &  &  & $( 2 ,  1 )^ 2_{9}$ & $38$ &  
 &  &  & $( 6 ,  0 )_{300}$ & $0$ \\     	
 &  &  & $( 4 ,  1 )_{30}$ & $34$ &  
 &  & $( 6 ,  1 )_{1134}$ & $( 2 ,  1 )^ 2_{20}$ & $14$ \\     	
 &  &  & $( 6 ,  1 )_{42}$ & $28$ &  
 &  &  & $( 4 ,  2 )_{105}$ & $4$ \\     	
 &  &  & $( 0 ,  0 )_{1}$ & $42$ &  
 &  &  & $( 4 ,  1 )_{175}$ & $8$ \\     	
 &  &  & $( 4 ,  0 )_{10}$ & $36$ &  
 &  &  & $( 6 ,  1 )_{729}$ & $0$ \\     	
 &  & $( 10 ,  4 )_{594}$ & $( 8 ,  4 )_{81}$ & $10$ &  
 &  &  & $( 0 ,  0 )_{1}$ & $20$ \\     	
 &  &  & $( 10 ,  4 )_{198}$ & $0$ &  
 &  &  & $( 4 ,  0 )_{84}$ & $10$ \\     	
 &  &  & $( 6 ,  3 )_{49}$ & $26$ &  
 &  & $( 6 ,  2 )_{1050}$ & $( 4 ,  2 )_{105}$ & $8$ \\     	
 &  &  & $( 8 ,  3 )_{126}$ & $18$ &  
 &  &  & $( 6 ,  2 )_{735}$ & $0$ \\     	
 &  &  & $( 4 ,  2 )_{25}$ & $38$ &  
 &  &  & $( 2 ,  1 )_{20}$ & $18$ \\     	
 &  &  & $( 6 ,  2 )_{70}$ & $32$ &  
 &  &  & $( 4 ,  1 )_{175}$ & $12$ \\     	
 &  &  & $( 2 ,  1 )_{9}$ & $46$ &  
 &  &  & $( 2 ,  0 )_{15}$ & $20$ \\     	
 &  &  & $( 4 ,  1 )_{30}$ & $42$ &  
 &  & $( 6 ,  3 )_{462}$ & $( 6 ,  3  )_{336}$ & $0$ \\     	
 &  &  & $( 2 ,  0 )_{6}$ & $48$ &  
 &  &  & $( 4 ,  2  )_{105}$ & $14$ \\     	
 &  & $( 10 ,  5 )_{286}$ & $( 10 ,  5  )_{121}$ & $0$ &  
 &  &  & $( 2 ,  1  )_{20}$ & $24$ \\     	
 &  &  & $( 8 ,  4  )_{81}$ & $20$ &  
 &  &  & $( 0 ,  0  )_{1}$ & $30$ \\     	
 &  &  & $( 6 ,  3  )_{49}$ & $36$ &  
 &  &  &  & \\     	
 &  &  & $( 4 ,  2  )_{25}$ & $48$ &  
 &  &  &  & \\     	 
 &  &  & $( 2 ,  1  )_{9}$ & $56$ &  
 &  &  &  & \\     	
 &  &  & $( 0 ,  0  )_{1}$ & $60$ &  
 &  &  &  &    	
\label{red} 
\end{longtable}

\newpage

\newpage

\newpage

\newpage

\newpage

\newpage

\newpage

\newpage

\newpage

\ed